\documentclass[galaxies,article,accept,moreauthors,pdftex]{Definitions/mdpi}

\firstpage{1}
\pubvolume{1}
\issuenum{1}
\articlenumber{0}
\pubyear{2026}
\copyrightyear{2026}
\externaleditor{Firstname Lastname} 
\datereceived{22 December 2025}
\daterevised{6 February 2026} 
\dateaccepted{14 February 2026}
\datepublished{ }
\hreflink{https://doi.org/} 

\usepackage{longtable}

\Title{The Optical Properties of Host Galaxies of Radio Sources in the Coma Cluster}

\Author{Xiaolan Hou $^{1}$\orcidA{}, Heng Yu $^{1,*}$\orcidB{}, Tong Pan $^{2}$\orcidC{}, Hu Zou $^{3,4}$\orcidD{}, Haoran Dou $^{1}$\orcidE{}, Emily Moravec $^{5}$\orcidF{} and Chengkui Li $^{6}$\orcidG{}}

\AuthorNames{Xiaolan Hou, Heng Yu, Tong Pan, Hu Zou, Haoran Dou, Emily Moravec, Chengkui Li}

\address{%
$^{1}$ \quad School of Physics and Astronomy, Beijing Normal University, Beijing 100875, China; \linebreak houxl@mail.bnu.edu.cn 
 (X.H.); 202321160002@mail.bnu.edu.cn (H.D.)\\
$^{2}$ \quad Leiden Observatory, Leiden University, Einsteinweg 55, 2333 CC 
Leiden, The Netherlands; tpan@strw.leidenuniv.nl\\
$^{3}$ \quad National Astronomical Observatories, Chinese Academy of Sciences, Beijing 100101, China; \linebreak zouhu@nao.cas.cn \\
$^{4}$ \quad School of Astronomy and Space Science, University of Chinese
Academy of Sciences, Beijing 101408, China \\
$^{5}$ \quad Green Bank Observatory, P.O. Box 2, Green Bank, WV 24944, USA; emoravec@nrao.edu \\
$^{6}$ \quad Key Laboratory of Particle Astrophysics, Institute of High Energy Physics, Chinese Academy of Sciences, Beijing 100049, China; lick@ihep.ac.cn}

\corres{Correspondence: yuheng@bnu.edu.cn}

\abstract{We present a comprehensive study of host galaxies of radio sources within the 1.35$R_{200}$ of the Coma cluster by combining deep $144\,\mathrm{MHz}$ observations from the LOFAR Two-Metre Sky Survey (LoTSS-DR2) with optical spectroscopy and photometry from DESI and SDSS. We identify 79 spectroscopically confirmed cluster members with reliable radio emission and classify them into compact, extended, and tailed subsamples according to their radio morphologies. By combining their radio and optical properties, we find compact radio sources are predominantly associated with massive, quiescent galaxies driven by $\mathrm{AGN}$ activity, while tailed sources are largely hosted by star-forming galaxies, tracing ongoing ram pressure stripping ($\mathrm{RPS}$). Using phase-space analysis and a projected infall time proxy ($d_R$), we find that extended sources are preferentially located in the cluster outskirts ($d_R > 1$), while tailed sources are concentrated in the intermediate infall region ($0.4 < d_R < 1.0$), highlighting the influence of the dense intracluster medium.
}


\begin{document}

\section{Introduction}

Galaxy clusters provide a powerful laboratory for studying how dense environments influence galaxy evolution. As~the largest gravitationally bound systems in the Universe, clusters host a range of environmental mechanisms, such as ram pressure stripping, tidal interactions (including galaxy harassment), and~starvation. These mechanisms act to remove, heat, or~redistribute the cold gas that fuels star formation. Consequently, these processes can substantially alter the star formation rates (SFRs), optical colors, and~morphological structures of galaxies, contributing to the well-established transformation from blue, star-forming spirals to red, quiescent early-type systems (e.g.,
 \cite{1972ApJ...176....1G,2006PASP..118..517B}). Understanding how these mechanisms operate across different galaxy populations is central to disentangling the environmental drivers of galaxy~evolution.

Radio continuum observations offer a complementary and uniquely sensitive probe of these evolutionary processes. Radio emission in galaxies primarily traces synchrotron radiation from relativistic electrons in magnetic fields, which are powered either by recent star formation (via cosmic rays from supernovae) or by active galactic nuclei (AGN) that launch jets and lobes (e.g., \cite{Bregman1990,Turner1994,Klein2018,Panessa2019}). Since radio emission is not affected by dust obscuration, it reveals activity that may otherwise remain hidden in optical surveys. Furthermore, the~diverse radio morphologies, including compact cores, extended jets, head-tail structures, and~diffuse disk emission, encode critical information about both internal energy sources and the interaction between the interstellar medium (ISM) of a galaxy and the intracluster medium (ICM). Recent studies utilizing LOFAR data have explicitly demonstrated that the dynamic environment of merging clusters significantly impacts these morphologies, leading to a higher population proportion of disturbed and bent sources compared to relaxed clusters \citep{2025ApJ...983..138R}. This makes radio data particularly valuable for characterizing environmental~processing.

The optical properties of radio-hosting galaxies in clusters provide additional insight into these mechanisms (e.g., \cite{Ledlow1996,Wing2011,Weeren2019}). Massive early-type galaxies typically exhibit red colors and low star formation rates, with~their radio emission predominantly powered by AGN activity (e.g., \cite{Salim2010,Capetti2022}). In~contrast, lower-mass or morphologically disturbed galaxies often display bluer colors, where radio emission is linked to ongoing or recently triggered star formation. In~these systems, the~star formation activity is sometimes enhanced or spatially reshaped by ram-pressure interactions (e.g., \cite{Best2004,Comerford2020}). Thus, the~combination of radio and optical diagnostics enables a more holistic view of the interplay between star formation, AGN activity, and~environmental~transformation.

\textls[-15]{The Coma Cluster (Abell 1656) is one of the nearest and richest galaxy clusters ($z \approx 0.023$), }{with characteristic radii of $R_{500} \sim$1.3 Mpc ($\approx 47^\prime$) and $R_{200} \sim$2 Mpc ($\approx 70^\prime$) \citep{2023A&A...670A.156C}.} {$R_{500}$ and $R_{200}$ are defined as the radii within which the average cluster density is 500 and 200 times the critical density of the Universe at the cluster's redshift, respectively.}
Due to its proximity and richness, Coma serves as an ideal laboratory for investigating the impact of dense environments on galaxy evolution and has been extensively observed across multiple wavelengths. It possesses deep optical photometric and spectroscopic coverage \cite{2019AJ....157..168D, 2025arXiv250314745D}, full-field low-frequency radio imaging from LOFAR, and~the latest X-ray measurements from the extended ROentgen Survey with an Imaging Telescope Array [eROSITA]; 
\cite{Churazov2021}. Together, these datasets provide a uniquely comprehensive, multi-band view of the cluster, enabling detailed studies of how the cluster environment shapes the properties and evolution of its member~galaxies.

\textls[-15]{Previous high-frequency radio observations with the VLA have already revealed the diverse nature of radio emission in Coma. Using deep 1.4 GHz imaging, \citet{Miller2009}} detected over 600 radio sources across two $30^\prime \times 50^\prime$ fields targeting the cluster core \mbox{($\sim0.4R_{200}$)} and the southwest infall region ($\sim1.25R_{200}$){, identifying 38 as confirmed cluster members}. These observations demonstrated that massive elliptical galaxies typically host AGN-powered emission, whereas fainter blue galaxies exhibit star formation-driven radio activity. {More recently, \citet{Lal2022} utilized the upgraded Giant Metrewave Radio Telescope (uGMRT) to provide a broader census of the cluster at 250--850 MHz across 4 deg$^2$ fields, characterizing the radio properties of 24 members and highlighting the presence of asymmetric or tail-like radio morphologies. Such features are consistent with ram-pressure interactions with the ICM during galaxy infall.}
These findings have established Coma as a benchmark system for studying environmental influences on galaxy activity through radio~observations.

{Compared to high/mid-frequency surveys, low-frequency observations at 144 MHz are significantly more sensitive to steep-spectrum and aged synchrotron emission. Such emission is expected to be prevalent in cluster galaxies undergoing environmental processes such as ram-pressure stripping.
The excellent surface-brightness sensitivity of the LOFAR Two-metre Sky Survey [LoTSS]; \citep{Shimwell2017} therefore makes it particularly well suited for detecting extended radio tails and diffuse emission associated with environmental transformation, complementing earlier 1.4 GHz studies.}

In this work, we combine deep LOFAR observations from the second data release of the LoTSS [LoTSS DR2]; \citep{2022A&A...659A...1S} with optical imaging from the DESI Legacy Imaging Survey (DR9) and spectroscopic measurements from DESI DR1. We utilize these data to conduct a systematic analysis of the radio activity and optical properties of member galaxies extending from the cluster core out to the periphery. The~primary goal of this study is to investigate the impact of the dense cluster environment on {radio emission of galaxies} across a wide range of local~densities.

This paper is organized as follows. In~Section~\ref{sec:data}, we describe the optical (DESI) and radio (LoTSS DR2) datasets, and~outline the construction of the spectroscopically confirmed Coma galaxy sample alongside their LOFAR radio counterparts. Section~\ref{sec:results} presents the physical properties of the host galaxies of radio sources in the cluster. In~Section~\ref{sec:env}, we examine the environmental dependence of these radio populations using the $\mathrm{R-V}$ diagram and the infall time proxy ($d_R$). Finally, Section~\ref{sec:conclusions} summarizes our main conclusions regarding the environmental modulation of radio activity in the Coma~cluster.

\section{Data and~Samples} \label{sec:data}
\subsection{Optical~Data} \label{ssec:optdata}
The optical photometric data used in this study are drawn from the DESI Legacy Imaging Survey (DR9; \cite{2019AJ....157..168D}), while the spectroscopic data are obtained from the DESI survey (DR1; \cite{2025arXiv250314745D}). We define a $4^\circ \times 4^\circ$ field ($\approx 6.72\,\mathrm{Mpc} \times 6.72\,\mathrm{Mpc}$ at $z = 0.023$) centered on NGC 4874 (R.A. = 194.90$^\circ$, Decl. = 27.96$^\circ$), the~Brightest Cluster Galaxy (BCG) of the Coma cluster. This field of view covers both the cluster core and its extended~outskirts.

From DESI DR1, we retrieved 22,360 galaxies with spectroscopic redshifts. Since the DESI target selection algorithm excludes very bright galaxies, we supplemented this dataset with spectroscopic measurements from the Sloan Digital Sky Survey (SDSS) [DR18] \citep{2023ApJS..267...44A} to recover bright sources. The~merged DESI+SDSS catalog contains 22,668 galaxies with reliable spectroscopic~redshifts.

\begin{figure}[H]
   \includegraphics[width=0.8\linewidth]{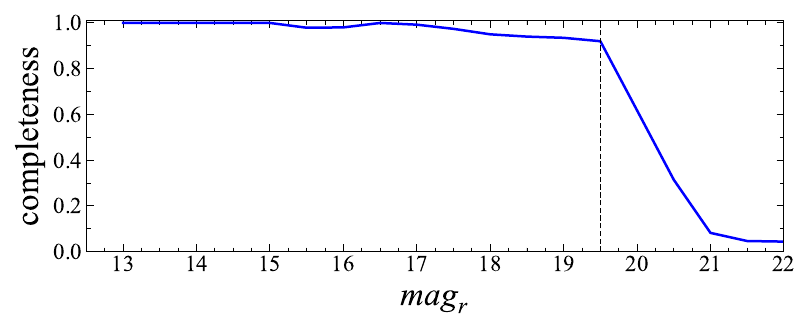}
   \includegraphics[width=0.8\linewidth]{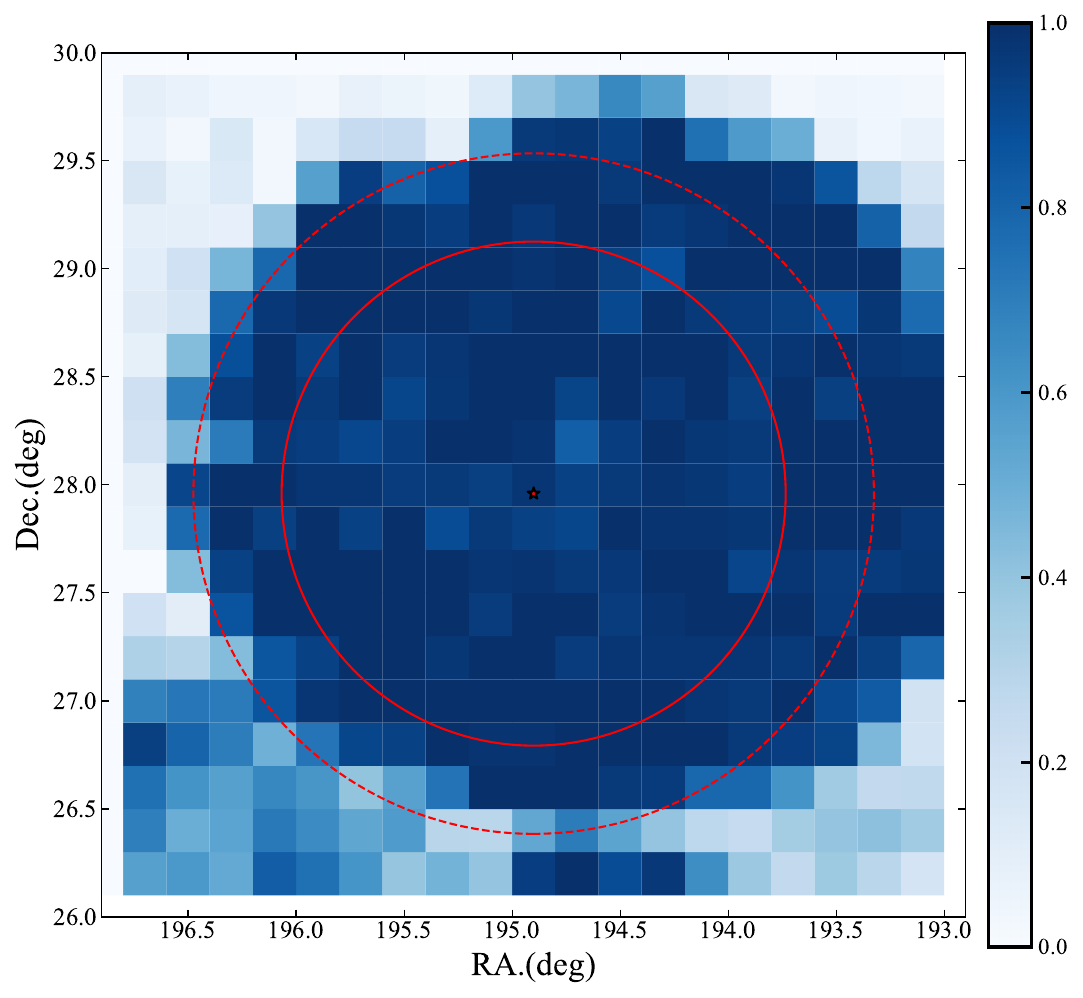}
   \caption{Spectroscopic data completeness. (\textbf{Upper panel}
): Spectroscopic completeness as a function of \textit{r}-band magnitude with in 1.35$R_{200}$. The~vertical dashed line denotes $r = 19.5$ mag. (\textbf{Bottom panel}): Spatial distribution of the spectroscopic completeness. The~colorbar represents the completeness at $r < 19.5$. The~red solid circle indicates the radius of $R_{200}$, and~red dashed circle indicates 1.35$R_{200}$.}
    \label{fig:comp}
\end{figure}

To assess the sample completeness, we selected 144,881 photometric sources with the $r$-band magnitude $mag_r < 22$ from the DESI Legacy Survey (DR9) within the defined field. The~mag$_r$ was derived from the measured flux according to the following relation: $\rm{mag}_r = 22.5 - 2.5 \log_{10} \left( \mathrm{f_r} / \mathrm{mw\_trans} \right)$. Here, the~f$_r$ is in units of nanomaggies and is corrected for the Galactic extinction using the $\mathrm{mw\_trans}$ factor provided by the survey. Using this photometric catalog, we evaluated the spectroscopic completeness as shown in Figure~\ref{fig:comp}. {The DESI Bright Galaxy Survey (BGS) consists of a flux-limited BGS Bright sample ($mag_r < 19.5$) and a color-selected BGS Faint sample ($19.5 < mag_r < 20.175$) \citep{Hahn2023}. Consistent with this selection, we find that the spectroscopic completeness exceeds 92\% for galaxies brighter than $mag_r = 19.5$, whereas it drops rapidly at fainter magnitudes.} 
The spatial completeness of bright galaxies $mag_r < 19.5$ remains above $\sim$94\% within a radius of $1.35R_{200}$ ($94.5^\prime$). Consequently, we defined a parent sample of 7957 
 galaxies with $mag_r < 19.5$ located within this central $1.35R_{200}$ circular region (covering an area of 7.8~deg$^2$) for cluster membership~determination.

\begin{figure}[H]
\includegraphics[width=0.8\linewidth]{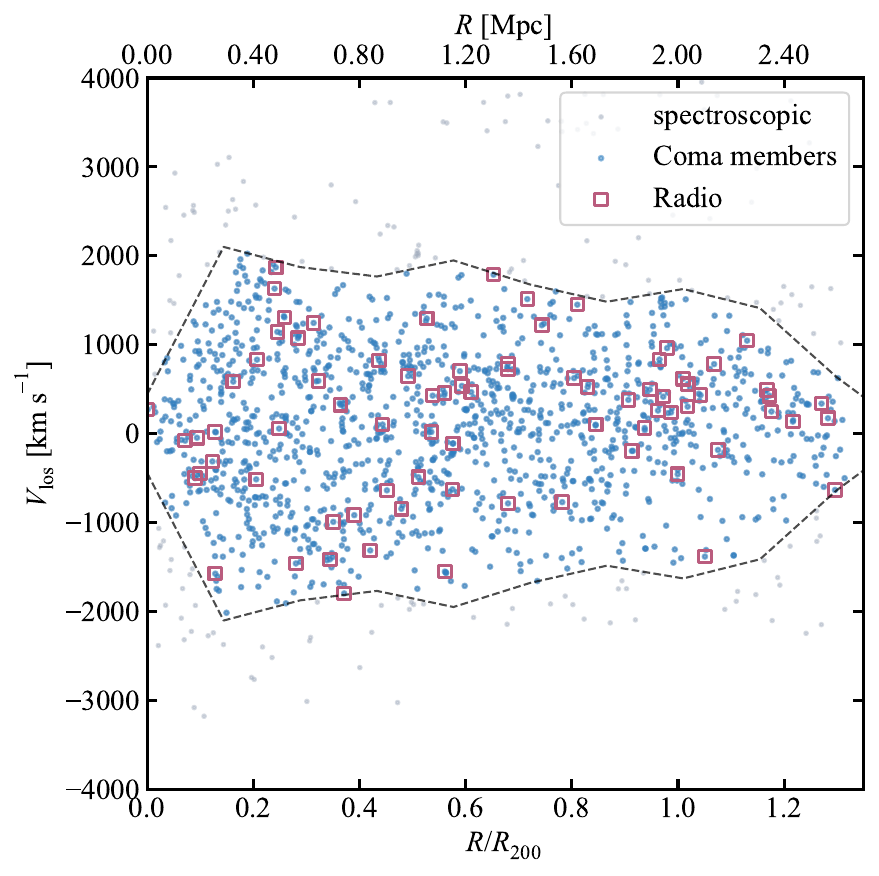}
\caption{The $R-V$ diagram ($R/R_{200}$ vs. $V_{los}$) of galaxies in the Coma Cluster field. Grey points represent spectroscopic galaxies with $r < 19.5$ mag. The~black dashed curves indicate the caustic lines. Blue points denote confirmed Coma cluster members located within the caustic boundaries{. The~79 galaxies with detected radio emission (LOFAR 144 MHz) are highlighted by red squares.}}
\label{fig:caustic}
\end{figure}

Cluster members were identified using the caustic technique \citep{Serra2011, Serra2013}.
{This method identifies cluster members by determining the boundary of the galaxy distribution in the $R-V$ diagram, or~the projected phase space (Figure \ref{fig:caustic}), where $R$ is the projected distance from the cluster center normalized with the $R_{200}$ and $V$ is the line-of-sight velocity relative to the cluster center. The~boundaries, or~'caustics', represent the escape velocity profile of the cluster as a function of radius. Galaxies located within the caustic envelopes are considered gravitationally bound members.}
This method yields 1329 Coma members with a line-of-sight velocity dispersion ($\sigma_{\mathrm{los}}$) of $818\,\mathrm{km~s^{-1}}$. Among~these members, 1258 galaxies have stellar mass {($M_\star$)} and star formation rate (SFR) measurements available in the DESI Stellar Mass and Emission Line Catalog \citep{2024ApJ...961..173Z},
where both quantities were derived via spectral energy distribution (SED) fitting using the \textsc{Cigale} code \citep{2019A&A...622A.103B}.

\unskip

\subsection{Radio~Sources}

\begin{figure}[H]
\scalebox{0.8}[0.72]{\includegraphics{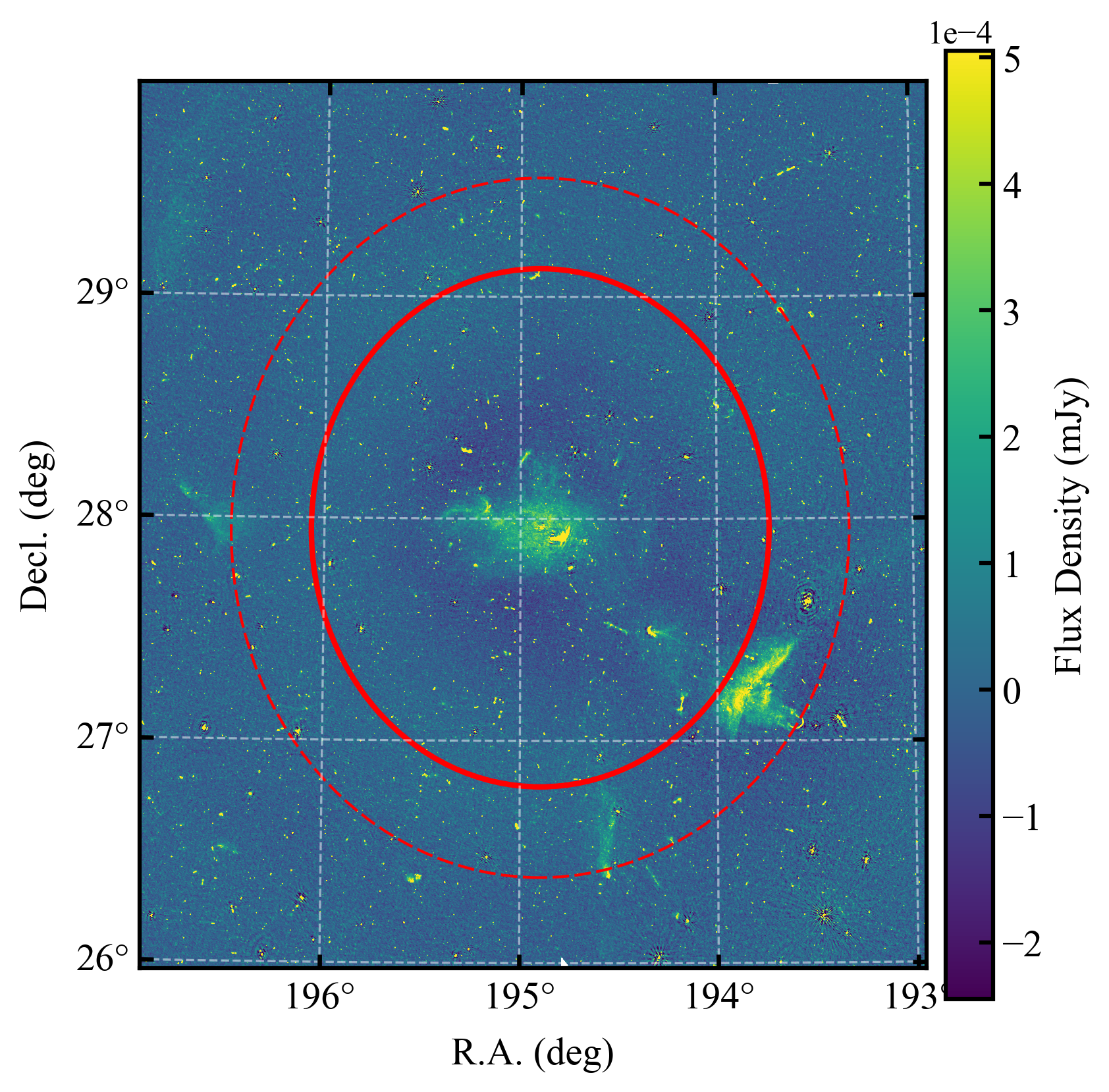}}
\caption{The LoTSS-DR2 radio continuum image of the Coma cluster region. This $4^\circ \times 4^\circ$ field is centered on NGC 4874, the~brightest central galaxy of the cluster. Red circles are the same as Figure~\ref{fig:comp}.}
\label{fig:radio}
\end{figure}

The radio continuum data for this study are drawn from {LoTSS-DR2 \citep{2022A&A...659A...1S}}. The~survey was conducted at a central frequency of $144\,\mathrm{MHz}$ with a resolution of $6^{\prime\prime}$. The~maps achieve a median $\mathrm{RMS}$ sensitivity of $83\,\upmu \mathrm{Jy~beam^{-1}}$ in Stokes I continuum. The~astrometric accuracy is $0.2^{\prime\prime}$, and~the flux density scale accuracy is approximately $10\%$. Based on peak brightness, the~point-source completeness is estimated to be $90\%$ at $0.8\,\mathrm{mJy~beam^{-1}}$. Within~the same $4^\circ \times 4^\circ$ region utilized for the optical sample, a~total of 11,287 radio sources were detected (Figure \ref{fig:radio}). 

\begin{figure}[H]
\includegraphics[width=0.8\linewidth]{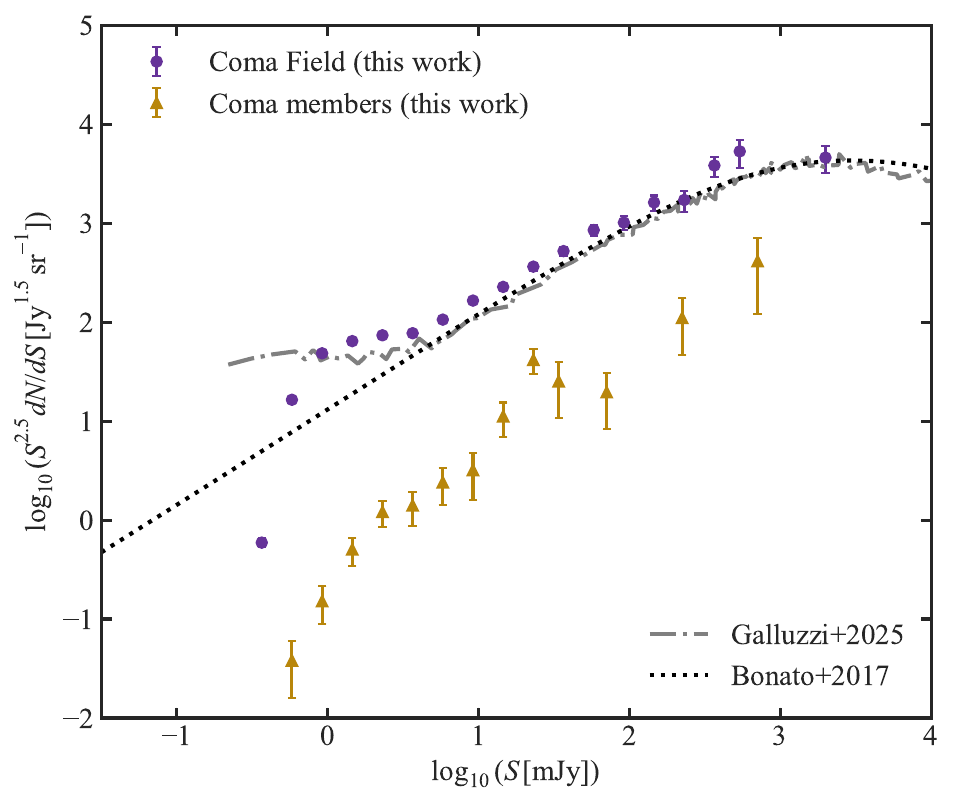}
\caption{\textls[-15]{Euclidean-normalized differential source counts at 144 MHz within $1.35R_{200}$. Data points show the total counts in the Coma field (purple circles) and confirmed cluster members (golden triangles). Overlaid are models from \citet{Galluzzi2025} (grey dash-dotted line) and \citet{Bonato2017} (black dotted line), accounting for diverse extragalactic radio populations. Error bars indicate $1\sigma$ uncertainties.}}
\label{fig:logNS}
\end{figure}

{To assess the statistical properties and completeness of our radio sample, we derived the Euclidean-normalized differential source counts (log$N$-log$S$) for galaxies within $1.35R_{200}$ of the cluster center (Figure \ref{fig:logNS}). The~observed distribution generally aligns with the evolutionary models of \citet{Bonato2017} and the empirical results from \citet{Galluzzi2025}. Notably, a~slight excess in the observed counts relative to the global models is apparent; this reflects the characteristic galaxy overdensity of the Coma cluster environment compared to the cosmic average.}

\unskip

To ensure the reliability of the radio detections, we established a threshold based on the Peak Signal-to-Noise Ratio (PNRs) calculated as the ratio of the peak flux density to the root-mean-square (RMS) noise within the source island, {with RMS values adopted from the LoTSS DR2 catalog}. We selected radio sources with PNR $> 4$ for subsequent analysis, resulting in a final sample of 10,354~sources.

We proceeded to identify the counterparts of these radio sources {that are} spectroscopically confirmed Coma member galaxies using an initial search radius of $6^{\prime\prime}$. There are 72~radio sources that meet the criteria. Recognizing that irregular radio/optical morphologies may lead to inaccurate centroid identification, we performed an additional visual examination for sources exhibiting positional offsets between $6^{\prime\prime}$ and {$20^{\prime\prime}$}. {There are 7 more members identified by the visual check.
Their optical/radio images are shown in Appendix \ref{a2}.}
After this verification process, our final sample comprises 79 Coma cluster members with radio emission {(see Figure~\ref{fig:caustic}).}

{We also compare our sample with 24 member sources reported by \citet{Lal2022}. There are 14 members overlapping, 8 sources not recognized as members by the caustic method. NGC 4827 and NGC 4869 are visible in the LOFAR image, but~not listed in the LoTSS DR2 catalog somehow. Thus they are not included in our sample. This would not affect the statistics of our sample.}

{Furthermore, the~source counts for the confirmed Coma cluster members are highlighted as golden triangles in Figure~\ref{fig:logNS}. The~distribution of these cluster members follows a similar trend to the total field counts, but~their numbers are too few to fully account for the observed excess, which may be due to gravitational lensing that amplifies the brightness of background radio sources.}

\begin{figure}[H]
    \includegraphics[width=0.28\linewidth]{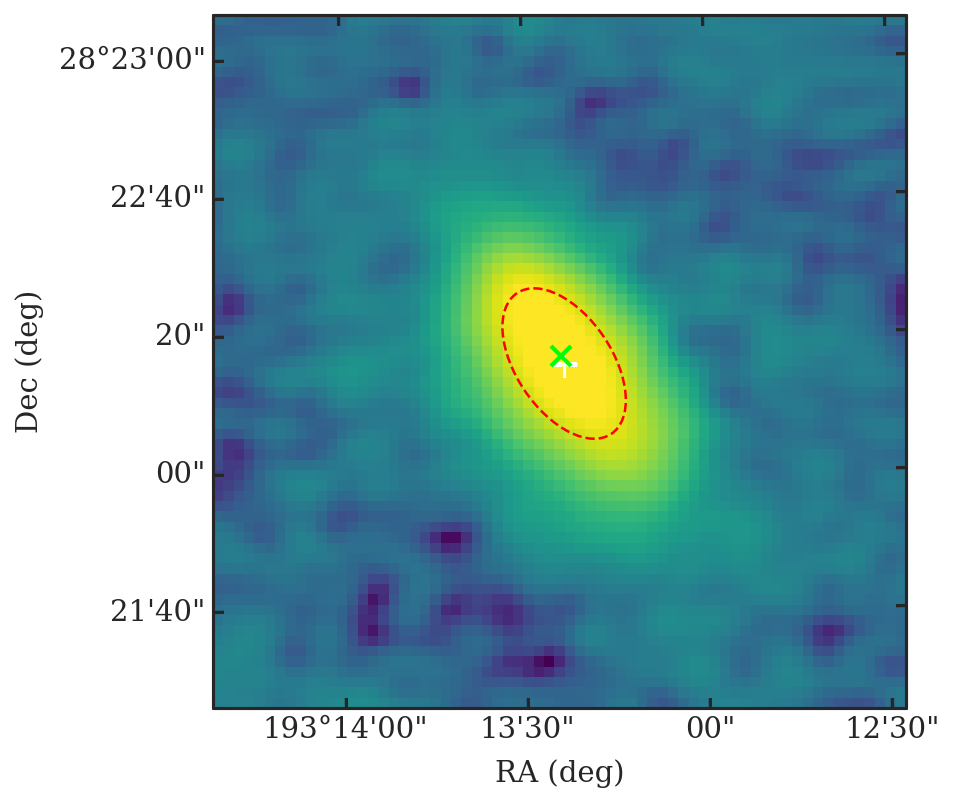}
    \includegraphics[width=0.28\linewidth]{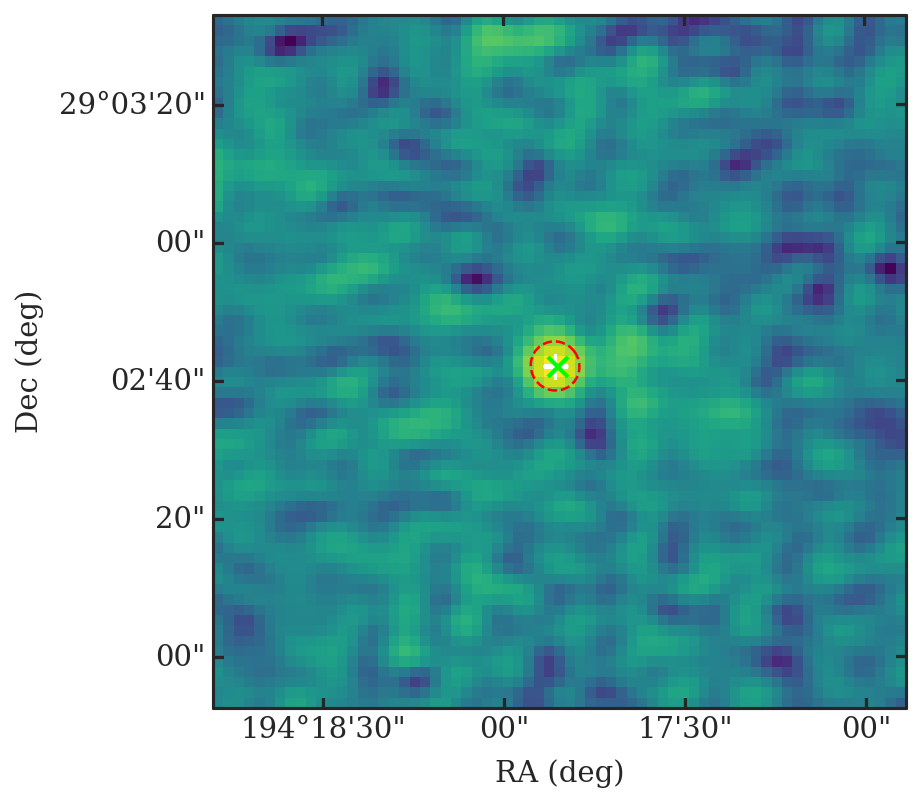}
    \includegraphics[width=0.28\linewidth]{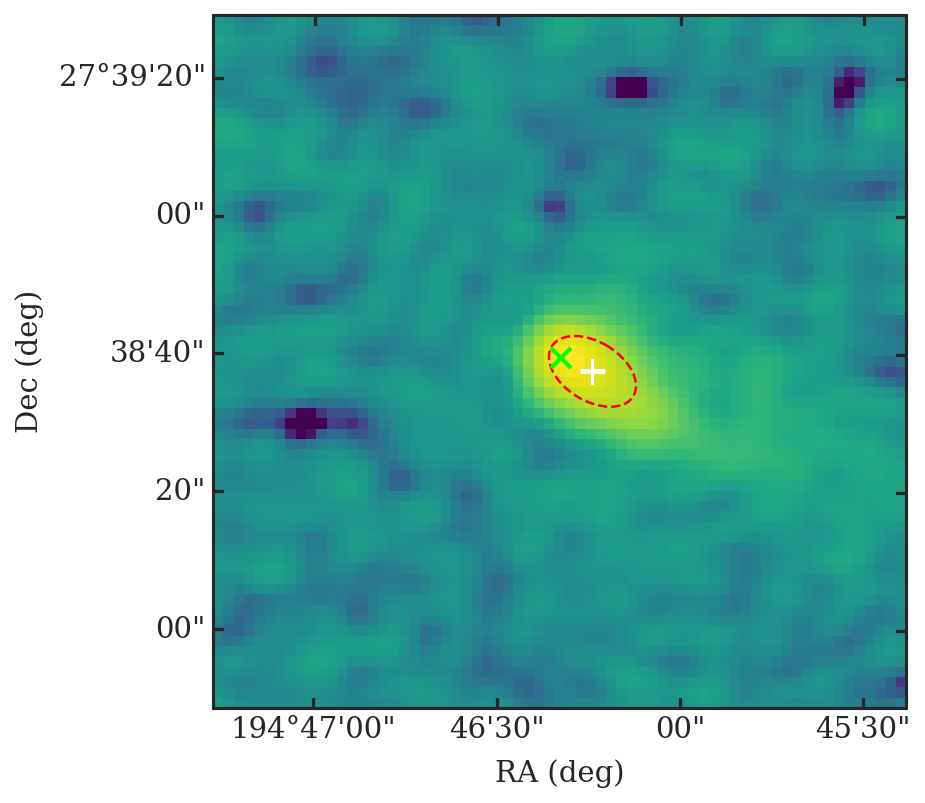}\\
    \includegraphics[width=0.28\linewidth]{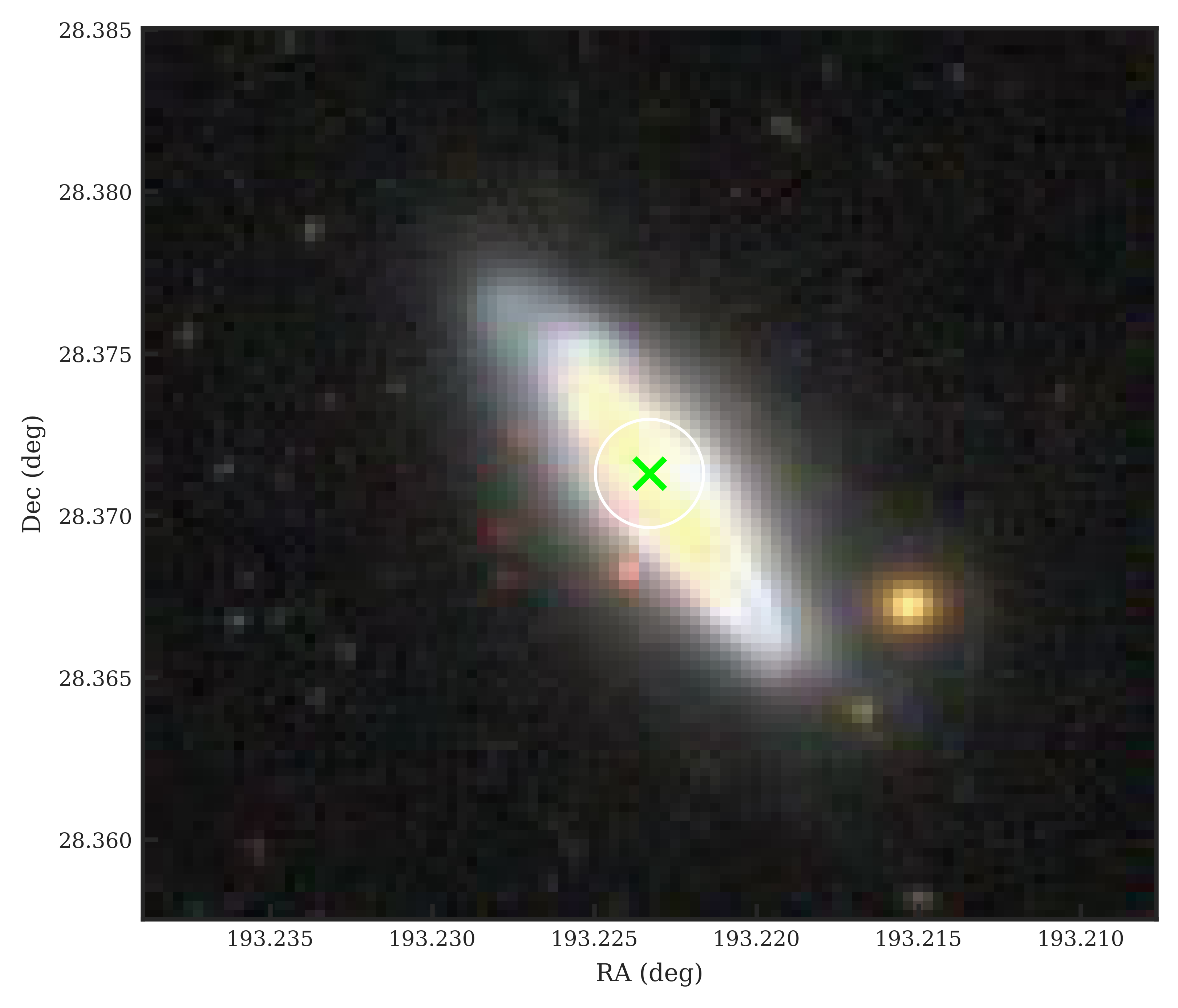}
    \includegraphics[width=0.28\linewidth]{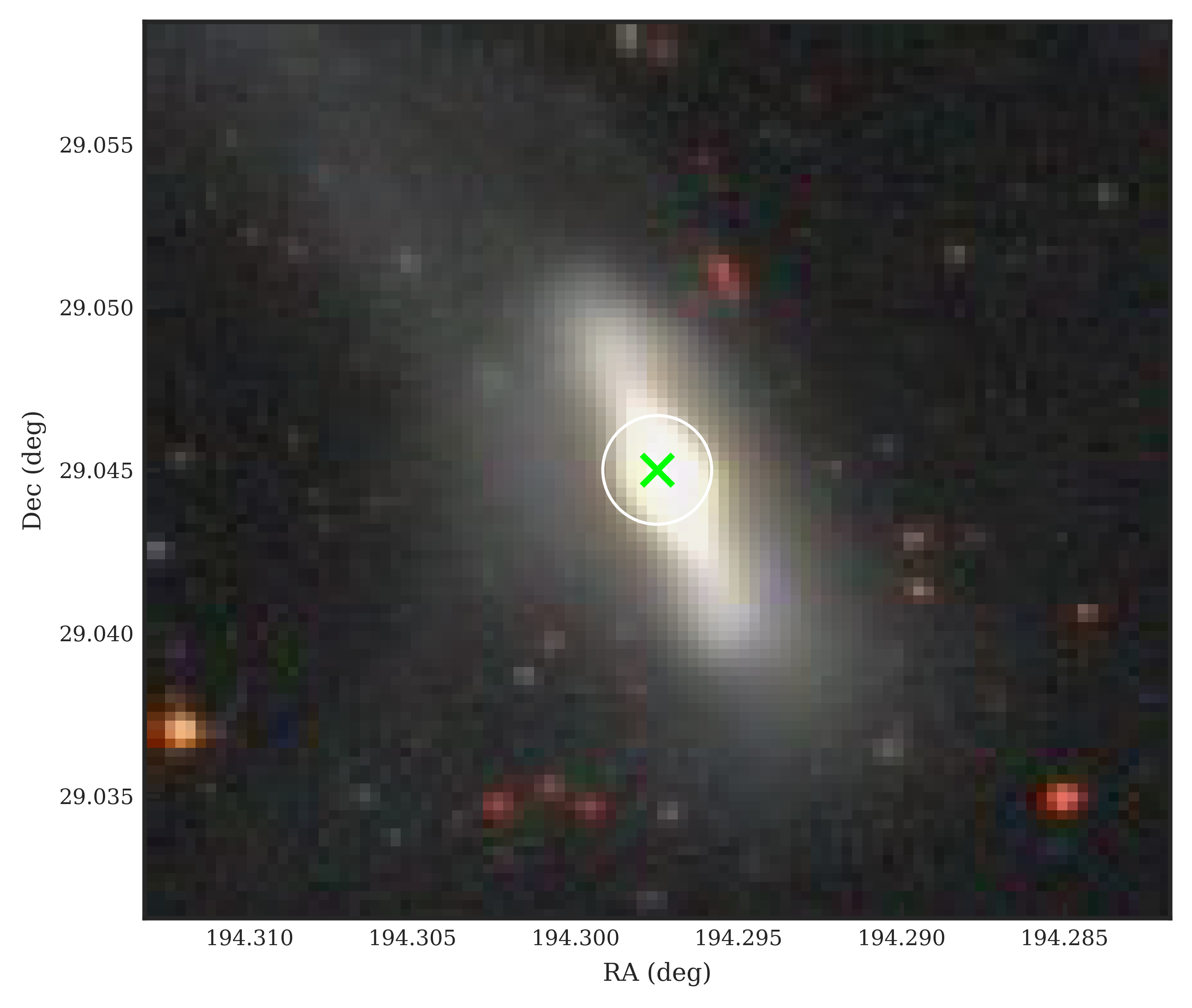}
    \includegraphics[width=0.28\linewidth]{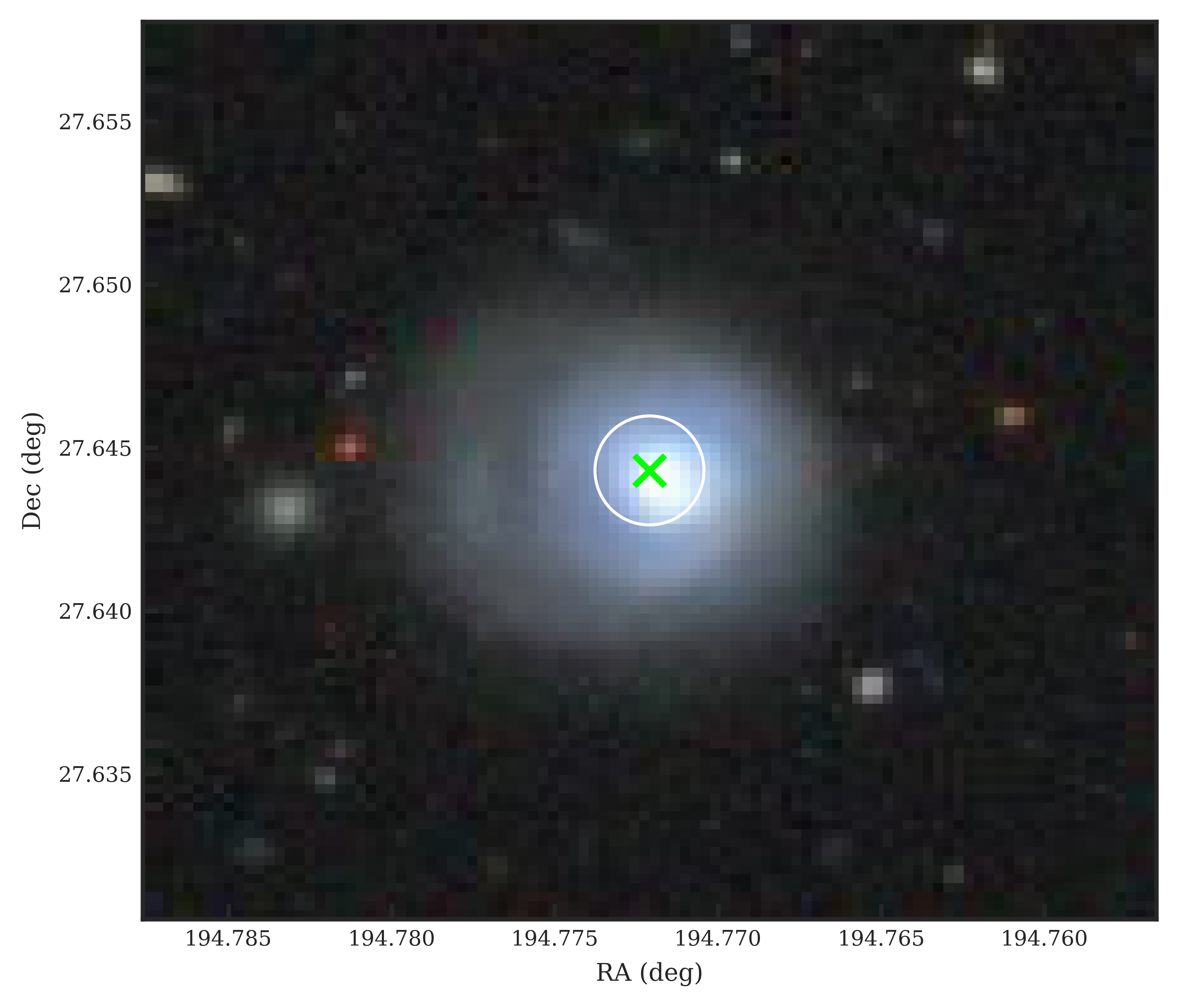}
  \caption{Examples 
  of radio sources classified by morphology: extended (E), compact (C), and~tailed (T). {The (\textbf{top panels}) display the LOFAR 144 MHz radio continuum images, each covering a field of view of 100$^{\prime\prime} \times 100^{\prime\prime}$, while the (\textbf{bottom panels}) show the corresponding optical images from DESI. In~all panels, green ``x'' symbols indicate the optical galactic centers, and~white crosses denote the centroids of the radio emission. The~red dashed ellipses in the radio images represent the morphology detections of the radio emission. The~white circles in the optical images have a radius of $6^{\prime\prime}$.}}
  \label{fig:samples}
\end{figure}

We classify these radio sources into three distinct types based on their radio morphology. Compact (C) sources are defined following the criteria in \citet{2025ApJ...983..138R}, exhibiting roughly circular emission with an angular size smaller than $12^{\prime\prime}$ (approximately twice the LoTSS survey resolution). Extended (E) sources show clearly elongated or irregular emission features with angular sizes greater than $12^{\prime\prime}$. We also isolated sources with distinct Tailed (T) structures as a separate category, typically characterized by a swept-back emission strongly indicative of galactic motion or ram-pressure stripping (RPS) processes. Exemplary sources of each classification are shown in Figure~\ref{fig:samples}.
{
Based on the visual inspection and morphological criteria described above, our 79 radio sources are classified into 31 compact sources, 25 extended sources, and~23 tailed structures.}
\vspace{-3pt}

\begin{figure}[H]
\includegraphics[width=0.66\linewidth]{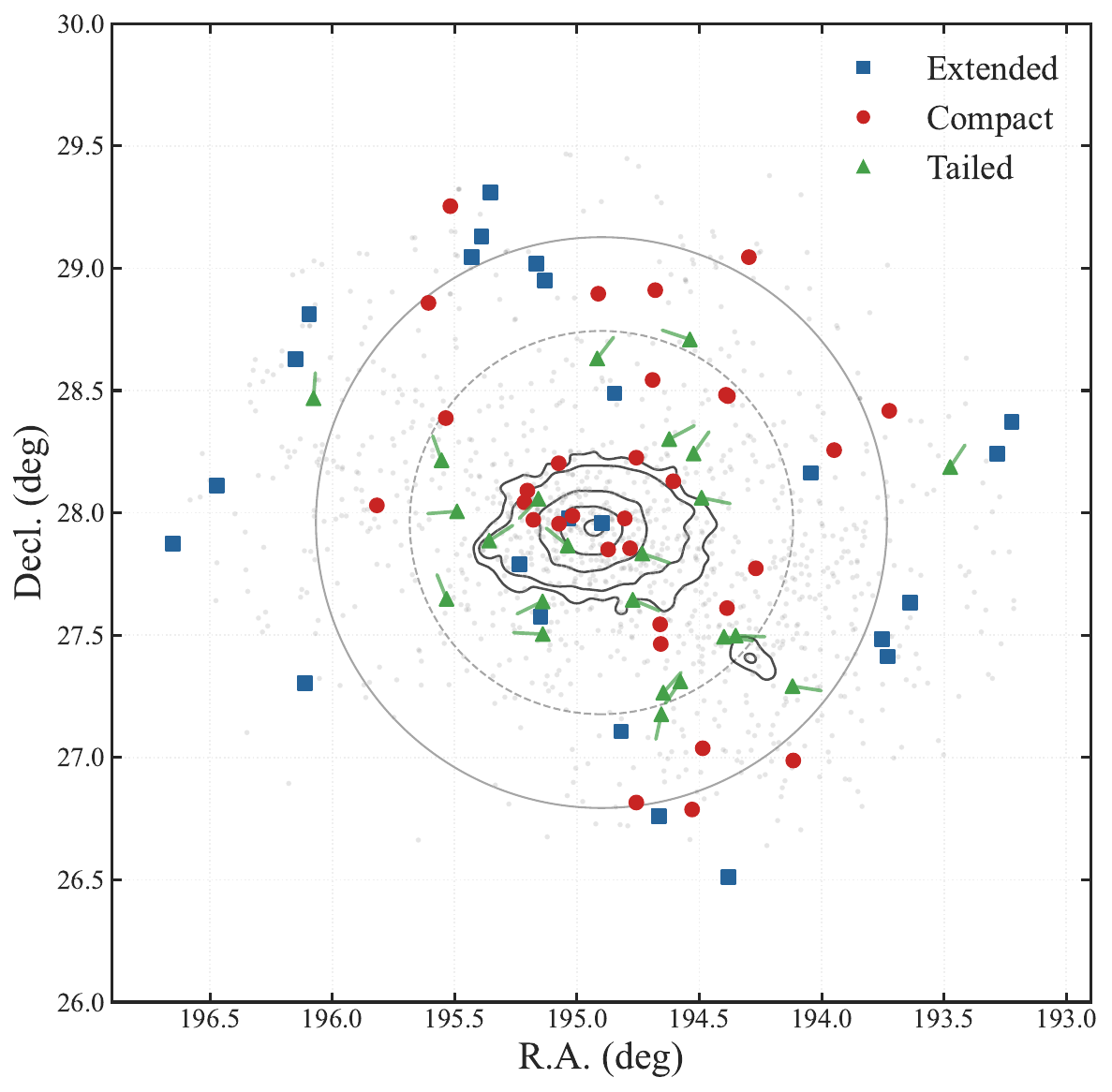}
\caption{The spatial distribution of 79 member radio sources. The~grey dots represent optical Coma members.
Radio sources are color-coded by morphology: blue squares for extended (E), red circles for compact (C), and~green triangles for tailed (T) structures. The~tails of tailed sources are marked with short lines. The~black contours show ROSAT X-ray surface brightness \citep{1995ApJ...439..113D}. The~solid grey circle represents the $R_{200}$, while the dashed circle represents $R_{500}$.}
\label{fig:radiosp}
\end{figure}

{
We also examined the spatial distribution of these 79 member radio sources and marked the tail directions of the tailed sources (Figure \ref{fig:radiosp}). It can be seen that they are primarily concentrated along the northeast--southwest direction, which aligns with the merger direction of the infalling NGC 4839 group. Most tailed sources are located within $R_{500}$, where the ICM is quite dense. However, we have not identified a clear or physically meaningful pattern in these tail directions; they are largely randomly distributed.}

\vspace{-3pt}
\unskip

\section{Optical Properties of Host~Galaxies} \label{sec:results}
\unskip

\subsection{Star-Forming and Quiescent~Populations}
\begin{figure}[H]
\includegraphics[width=0.66\linewidth]{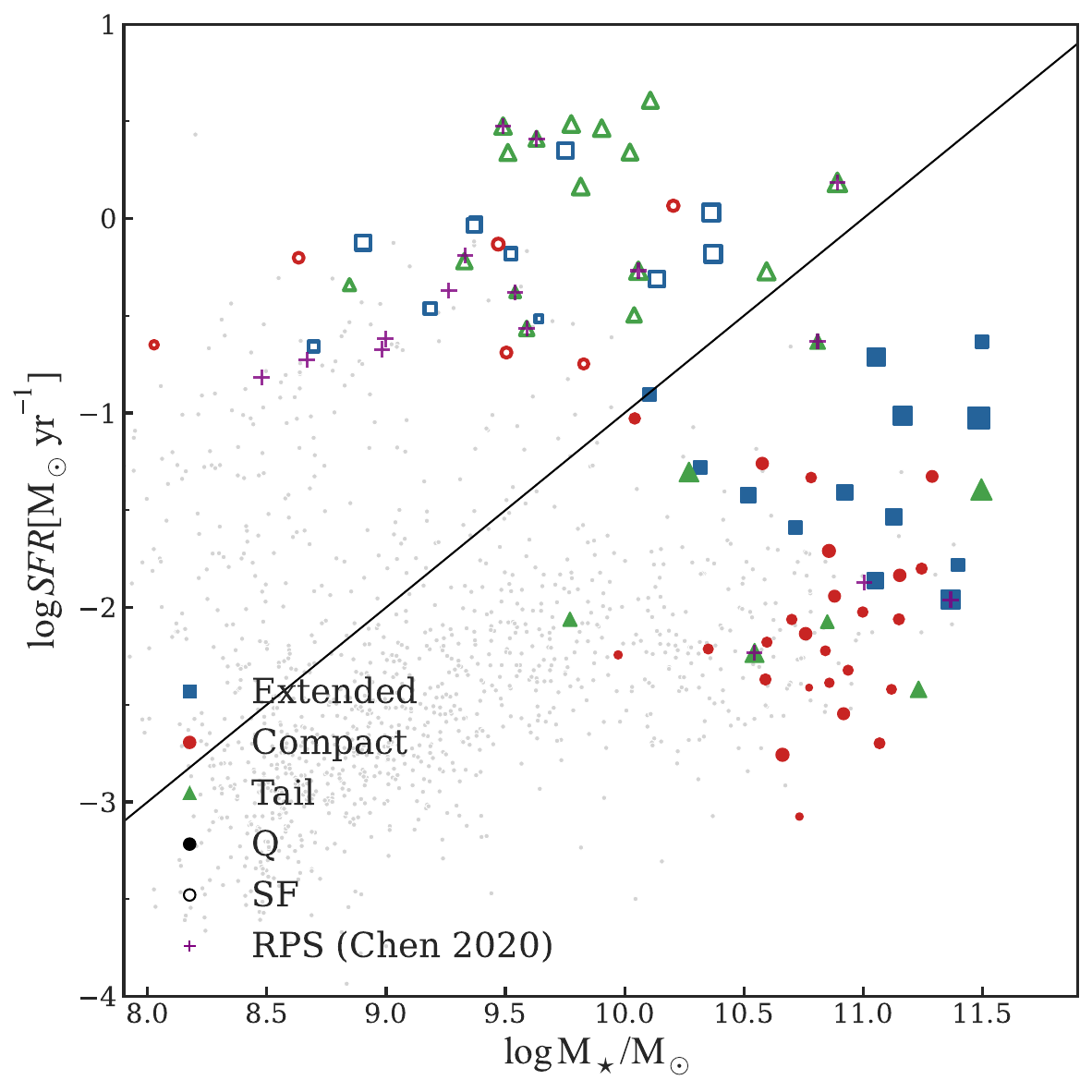}
\caption{Distribution 
of galaxies in the $\log \mathrm{SFR} $-$\log \rm{M_\star} $ plane.  The~grey dots represent optical Coma members.
Radio sources are color-coded by morphology: blue squares for extended (E), red circles for compact (C), and~green triangles for tailed (T) structures. The~black solid line denotes a specific star formation rate of $\log \mathrm{sSFR} = -11 \, \mathrm{yr}^{-1}$. Galaxies are further categorized based on this threshold: star-forming galaxies (SF) are shown as open symbols, while quiescent galaxies (Q) are shown as filled symbols.  The~size of markers reflects the radio luminosity at 144 MHz ($\log L_{144}$). {The purple crosses represent optical RPS galaxies identified by \citet{2020MNRAS.496.4654C}.}}
\label{fig:m_sfr}
\end{figure}

To explore the optical properties of the host galaxies of radio sources in our sample, we adopt a specific star formation rate ($\mathrm{sSFR}$) threshold of $\log(\mathrm{sSFR}) = -11\,\mathrm{yr^{-1}}$ \citep{Wetzel2012} to distinguish between actively star-forming ($\mathrm{SF}$) and quiescent ($\mathrm{Q}$) systems.
Based on this criterion, our sample comprises 34 star-forming galaxies and 45 quiescent galaxies. Figure~\ref{fig:m_sfr} shows the distribution of these galaxies in the $\log(\mathrm{M_\star})$-$\log(\mathrm{SFR})$ plane. Extended (E) sources are represented by blue squares, Compact (C) sources by red circles, and~Tailed (T) structures by green triangles. Quiescent galaxies are indicated by filled markers, while star-forming galaxies are shown with open~markers.

The host galaxies in our sample clearly exhibit a bimodal distribution, populating either the actively star-forming sequence ({$\log \mathrm{SFR} > -1$}) or the region of high stellar mass ($\log \mathrm{M_\star} > 10.5\,\mathrm{M}_\odot$). A~strong correlation is evident between radio morphology and optical state: {the} vast majority (25 out of 31) of compact sources are associated with quiescent galaxies, suggesting that their radio emission is primarily attributed to the central $\mathrm{AGN}$ activity, which is a typical characteristic of massive elliptical systems \mbox{\citep{Fabian2012, Tadhunter2016}}. In~contrast, a~significant portion (16 out of 23) of the tailed structures originates from star-forming galaxies. Specifically, nine of these star-forming tailed sources have been previously confirmed as optical $\mathrm{RPS}$ galaxies by \citet{2020MNRAS.496.4654C}. Although~the RPS validation sample from \citet{2020MNRAS.496.4654C} only covers the central part of our field ($\sim1.25\,\mathrm{deg^2}$), this significant overlap indicates that the radio tails primarily originate from star formation within cool gas stripped by environmental processes, such as RPS or tidal interactions.
Conversely, the~radio tails observed in quiescent galaxies are more likely related to jets and outflows powered by their central $\mathrm{AGN}$. The~extended radio sources, however, show a near-equal division between star-forming galaxies (12) and quiescent galaxies (13). It is inherently difficult to morphologically disentangle whether the extended radio emission originates from star-forming regions (diffuse disk emission) or from $\mathrm{AGN}$-driven jets/lobes in this class. However, the~$\mathrm{SFR}$ measured in the optical band provides a crucial constraint to quantify the relative contribution from star formation activity. {Future studies incorporating far-infrared (FIR) flux would be beneficial in providing a more robust, dust-independent census of star formation and enabling improved cross-calibration with radio-based indicators.}
\vspace{-3pt}
\unskip

\subsection{AGN-Dominated Versus Star Formation-Dominated~Activity}

To further investigate whether the radio emission in these galaxies is dominated by $\mathrm{AGN}$ or star formation ($\mathrm{SF}$), we examined the relationship between the $144\,\mathrm{MHz}$ radio luminosity ($\mathrm{L_{144}}$) and the $\mathrm{SFR}$ of the galaxies. The~$\mathrm{L_{144}}$ was calculated for each galaxy from the observed total integrated flux density ($F_{\mathrm{tot}}$) reported in the LoTSS-DR2 catalog, following standard radio $\mathrm{K}$-correction procedures. The~luminosity was computed using the following cosmological relation \citep{Igo2024}:
\begin{equation}
{L_{144} [\text{W Hz}^{-1}] = 4 \pi D_{L}^{2} [\text{m}^2] \, F_{\text{tot}} [\text{mJy}]  10^{-29}  (1+z)^{\alpha - 1}}
\end{equation}

Here, $D_{L}$ is the luminosity distance in meters, $F_{\mathrm{tot}}$ is the observed total integrated flux density in $\mathrm{mJy}$, and~$z$ is the spectroscopic redshift. Following conventional practice for $\mathrm{K}$-correction in this band, the~radio spectral index $\alpha$ (defined as $S \propto \nu^{-\alpha}$) is fixed at a constant value of $\alpha = 0.7$ \citep{Condon1992}.
{At low redshift, the~choice of $\alpha$ has a minimal impact on the $K$-correction. For~the Coma cluster ($z \approx 0.023$), varying $\alpha$ from 0.7 to 0.3 results in a change of only $\sim 1.6\%$ in the derived luminosity.}

\begin{figure}[H]
\includegraphics[width=0.68\linewidth]{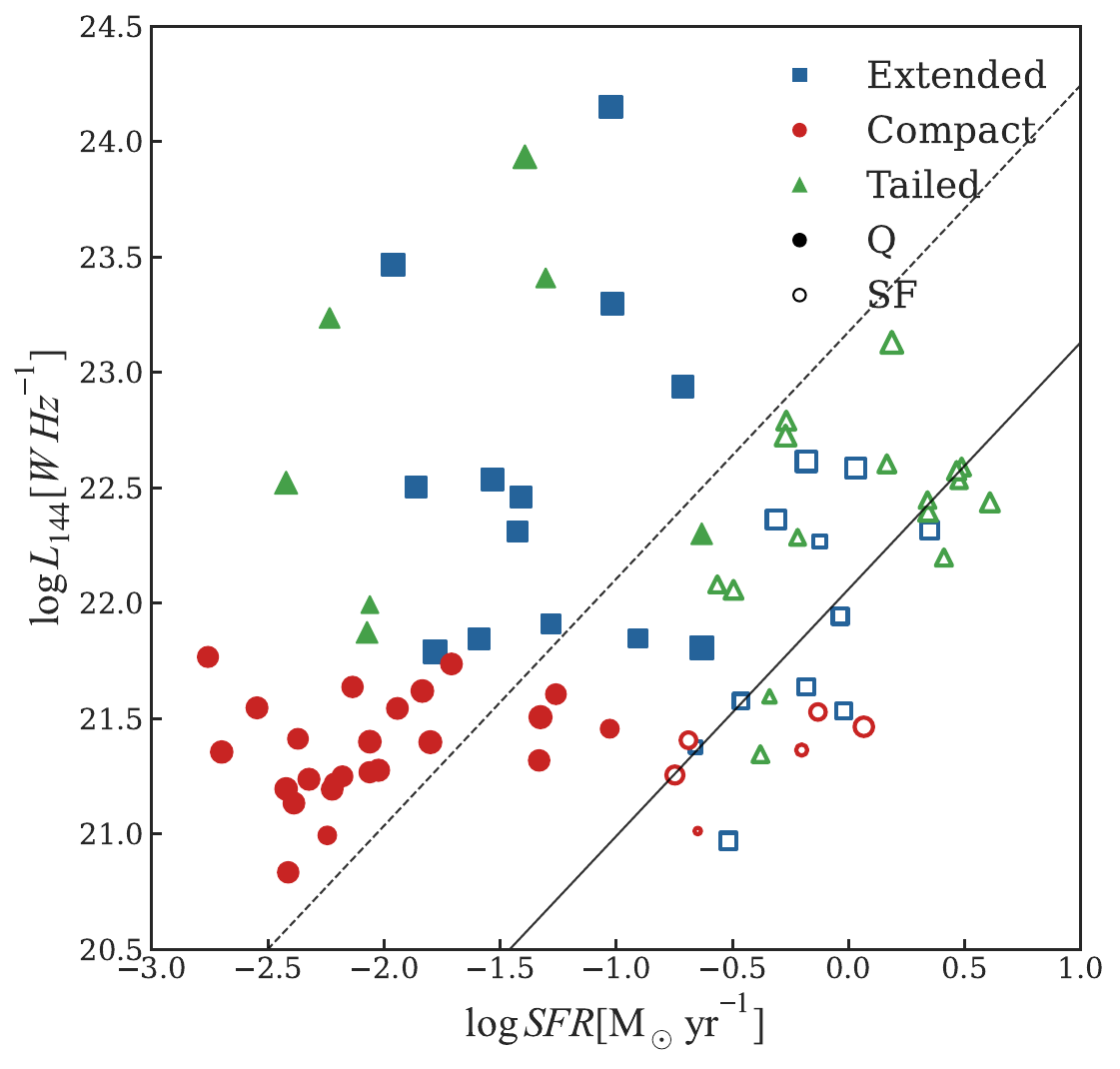}
\caption{SFR versus the radio luminosity of radio sources within the Coma members. The~black {solid} line represents the best-fit relation for star-forming galaxies derived by \citet{Gurkan2018}, {and the black dashed line represents $3\sigma$ above this relation}. Marker types and colors are the same as in Figure~\ref{fig:m_sfr}. The~size of markers is scaled with stellar masses ($M_\star$) of each~galaxy.}
\label{fig:radio_sfr}
\end{figure}

The distribution of $\mathrm{L_{144}}$ and $\log \mathrm{SFR}$ for Coma members is shown in Figure~\ref{fig:radio_sfr}. It can be seen that compact sources (C) are relatively dim, clustering towards the lower left portion of the figure, consistent with low-power $\mathrm{AGN}$ activity. Radio sources exhibiting extended structures (E) or tailed structures (T) are more diverse, displaying a range of luminosities and $\mathrm{SFR}$s.
The $6^{\prime\prime}$ resolution of the LoTSS survey corresponds to a physical scale of $2.8\,\mathrm{kpc}$ at the distance of the Coma cluster. This resolution provides crucial visual insight into the origin of the radio emission. Extended radio emission originating purely from star-forming regions typically shows a good morphological correspondence with the galaxy's optical~structure. 

To quantitatively disentangle the $\mathrm{AGN}$ and $\mathrm{SF}$ contributions, we utilize an empirical radio-$\mathrm{SFR}$ relation derived from the LoTSS Deep Fields \citep{Gurkan2018,Best2023}:
\begin{equation}
    \log_{10}L_{144}[\rm{W~Hz}^{-1}] = 22.06 + 1.07 \, \log_{10}SFR,
\end{equation}
This 
relation, depicted by the black dashed line in Figure~\ref{fig:radio_sfr}, \citet{Gurkan2018}, represents the expected radio luminosity contributed by star formation alone. {Radio sources with measured luminosities exceeding this relation by more than a factor of three are considered to require an additional contribution from an $\mathrm{AGN}$.}
We find that the star-forming galaxies (open markers) and quiescent galaxies (filled markers) maintain a distinct separation in this diagram. This observed separation is consistent with the expectations set by the radio--$\mathrm{SFR}$ relationship. Consequently, it is reasonable to conclude that the radio activity observed in quiescent galaxies is predominantly $\mathrm{AGN}$-dominated, while in $\mathrm{SF}$ galaxies, a~significant portion of their radio emission is contributed by ongoing star formation~activity.
\vspace{-3pt}
\unskip

\subsection{The Color--Magnitude~Relation}
\begin{figure}[H]
\includegraphics[width=0.53\linewidth]{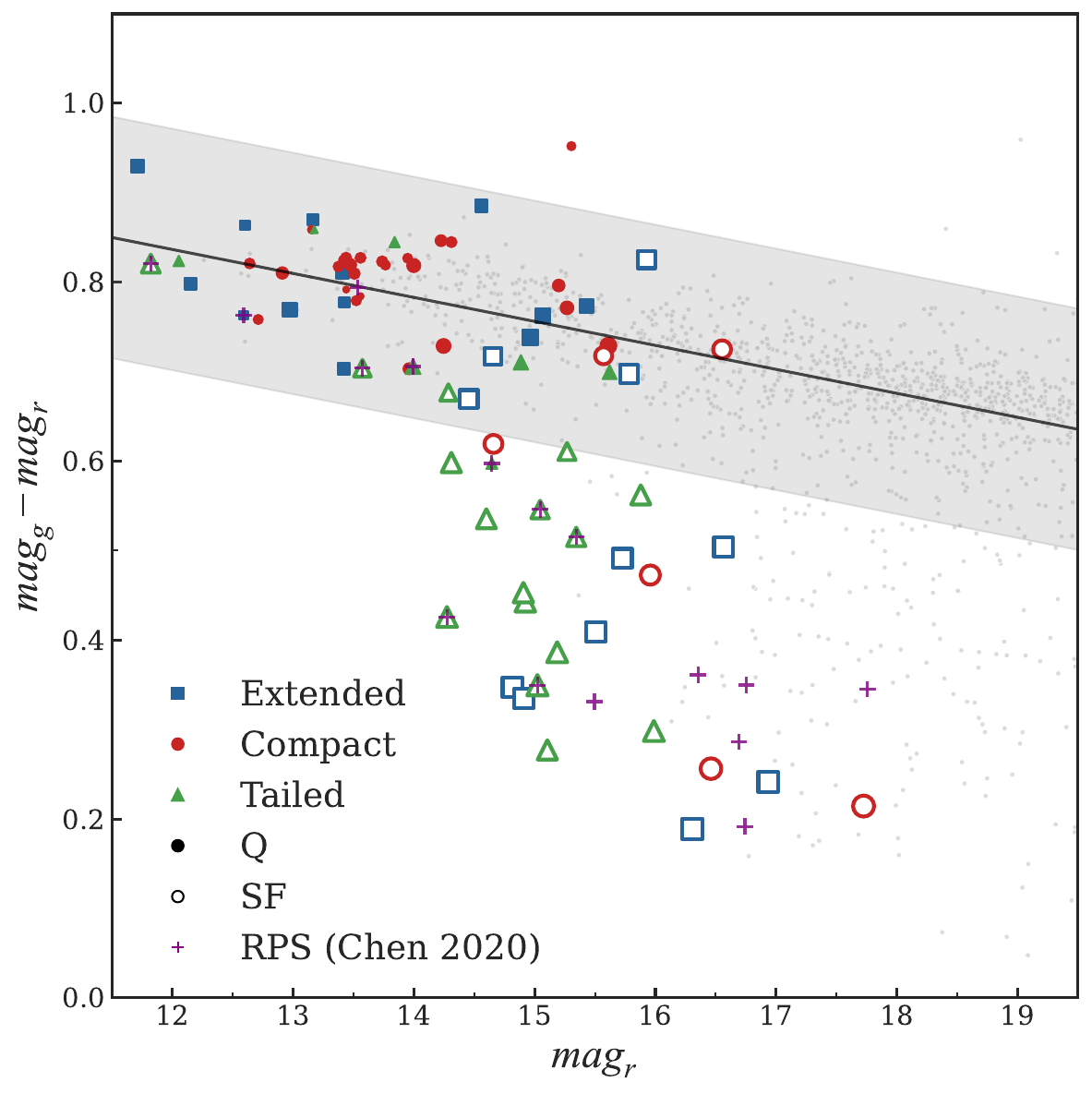}
\caption{Color (mag$_g$-mag$_r$) as a function of mag$_r$ for Coma cluster galaxies. Marker types and colors are the same as in Figure~\ref{fig:m_sfr}. The~size of markers is scaled with $\log \mathrm{sSFR}$. The~black solid line is the fitted red sequence derived from the color distribution of the spectroscopic data, with~the shaded grey area indicating the $3\sigma$ confidence interval of the~fit.}
\label{fig:m_color}
\end{figure}

We examined the distribution of the host galaxies of radio sources on the Color--Magnitude Diagram ($\mathrm{CMD}$), as~depicted in Figure~\ref{fig:m_color}. To~establish the baseline stellar population for comparison, we first determined the Color--Magnitude Relation ($\mathrm{CMR}$), commonly referred to as the Red Sequence ($\mathrm{RS}$). The~$\mathrm{RS}$ was derived using an iterative $\sigma$-clipping least-squares fitting procedure, performed exclusively on the spectroscopically confirmed Coma cluster members. This technique effectively suppresses contamination from blue cloud galaxies and interlopers, accurately capturing the intrinsic scatter of the passive galaxy population \citep{Gladders2000}. At~each iteration, a~linear function of the form $\rm{mag}_g-\rm{mag}_r = \alpha \rm{mag}_r + \beta$ was fitted to the data. Galaxies whose color residuals exceeded $3\sigma$ from the current best-fit relation were iteratively excluded, and~the fitting process was repeated until the slope ($\alpha$) and intercept ($\beta$) converged to within a relative change of $10^{-4}$, or~until a maximum of ten iterations was reached \citep{Sciarratta2019}. This procedure yielded a stable and robust characterization of the cluster $\mathrm{RS}$. The~final fitted $\mathrm{RS}$ is described by the linear relation: $\rm{mag}_g - \rm{mag}_r = -0.027 \rm{mag}_r + 1.158$.

Generally, the~radio sources in our sample are distributed primarily along the bright end of the $\mathrm{CMD}$. {The $\mathrm{AGN}$-dominated radio sources (filled markers in Figure~\ref{fig:m_color}), which are predominantly associated with quiescent galaxies identified via radio excess emission, are found to lie close to the red sequence.}
This observed behavior is consistent with the established picture in which radio-loud $\mathrm{AGN}$ preferentially reside in massive, quiescent early-type hosts whose cold gas reservoirs have been largely depleted or stabilized against further star formation via internal feedback mechanisms \citep{Best2004}.

\textls[-15]{In contrast, the~star-forming galaxies {(open markers in Figure~\ref{fig:m_color})} are generally bluer. This blue color directly indicates a high fraction of young stellar populations and ongoing star formation activity. This trend persists across the full mass range, implying that $\mathrm{RPS}$ candidate galaxies
do not simply occupy a particular stellar mass regime; rather, they represent a star-forming subset active at each mass scale. Their blue colors suggest that these systems are still actively forming stars, which may be maintained or temporarily enhanced during ram-pressure interaction as the external $\mathrm{ICM}$ pressure compresses the interstellar medium and triggers short-lived bursts of star formation before the eventual quenching process~begins.}
\vspace{-12pt}


\section{Environment~Influence} \label{sec:env}

\begin{figure}[H]
\includegraphics[width=0.8\textwidth]{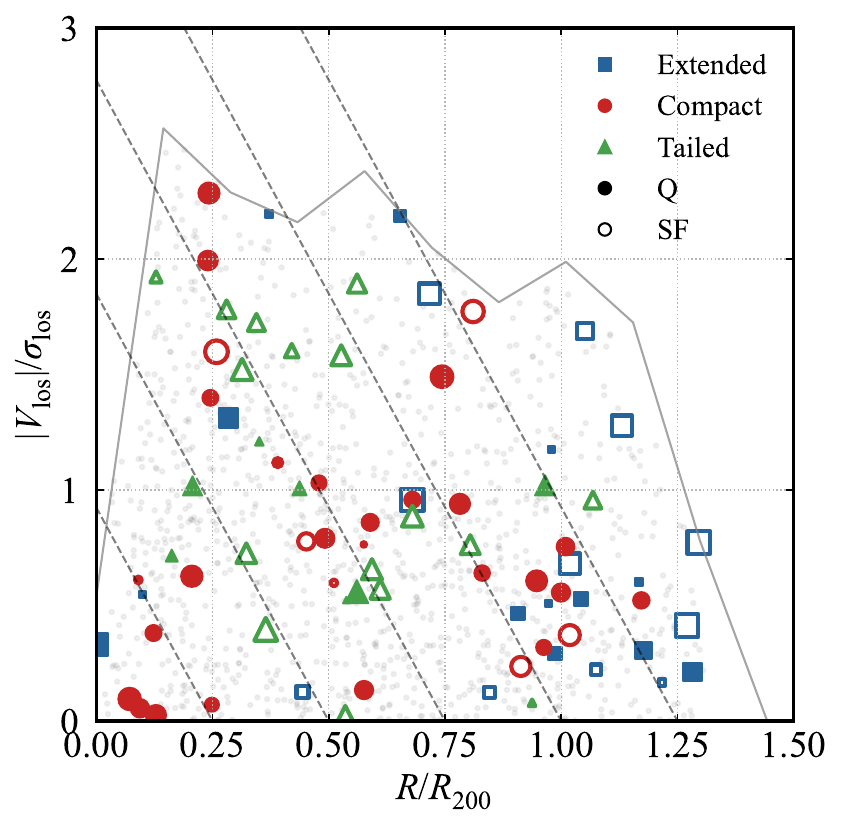}
\caption{$R$-$V$ diagram for galaxies {within} 1.35 $R_{200}$ around the center of Coma cluster. The~solid broken line shows the same caustic line as in Figure~\ref{fig:caustic}. The~parallel dashed lines represent characteristic dynamical zones \citep{2025A&A...699A..95D}.
Marker types and colors are the same as in Figure~\ref{fig:m_sfr}. The~size of markers is scaled with $\log L_{144}$.}
\label{fig:rv}
\end{figure}

To evaluate the environmental influences on the radio properties of our sample, we examined the distribution of radio sources in the $\mathrm{R-V}$ diagram (Figure\ref{fig:rv}), where $R$ is the projected cluster-centric radius and $V$ is the line-of-sight velocity. The~diagram plots the projected cluster-centric radius normalized by $R_{200}$ against the line-of-sight velocity offset normalized by the velocity dispersion ($\Delta V_{\mathrm{los}} / \sigma_{\mathrm{los}}$, where $\Delta V_{\mathrm{los}} = |V_{\mathrm{los}}|$). This representation provides valuable insight into the dynamical histories of galaxies within the cluster, as~different regions correspond to distinct stages of infall and environmental processing (e.g., \cite{Rhee2017,Pasquali2019,2025A&A...699A..95D}). To~quantify {where infalling galaxies are located within the cluster environment}, we adopt the method developed by \citet{2025A&A...699A..95D}, which uses a set of parallel oblique lines in the $\mathrm{R-V}$ diagram. {While $R_{200}$ serves as a global proxy for the clusters dynamical boundary, we also incorporate $R_{500}$ into our radial analysis. This choice is motivated by the strong correlation between $R_{500}$ and the hot ICM properties, facilitating a more direct comparison between radio morphologies and the ICM density peak.} We define the $\mathrm{x}$-intercept of these lines as $d_R$, which serves as {a relative proxy for the infall stage} of the cluster member galaxies. This $d_R$ distance is defined by $d_R = -(|V_{los}|/\sigma_{los} - kR/R_{200})/k$, where the slope $k=-3.7$ is adopted from the $\mathrm{TNG300}$ simulation \citep{2025A&A...699A..95D}. It can be directly compared to the normalized projected radius $R/R_{200}$. {By construction, smaller values of $d_R$ correspond to galaxies that have likely experienced earlier infall and are typically located closer to the cluster core, whereas larger values of $d_R$ indicate later infall stages and are generally associated with galaxies in the cluster outskirts. In~this framework, galaxies with $d_R > 1$ correspond to galaxies located beyond $R_{200}$ and~therefore trace the cluster outskirts.}
\vspace{-5pt}

\begin{figure}[H]
\includegraphics[width=0.8\textwidth]{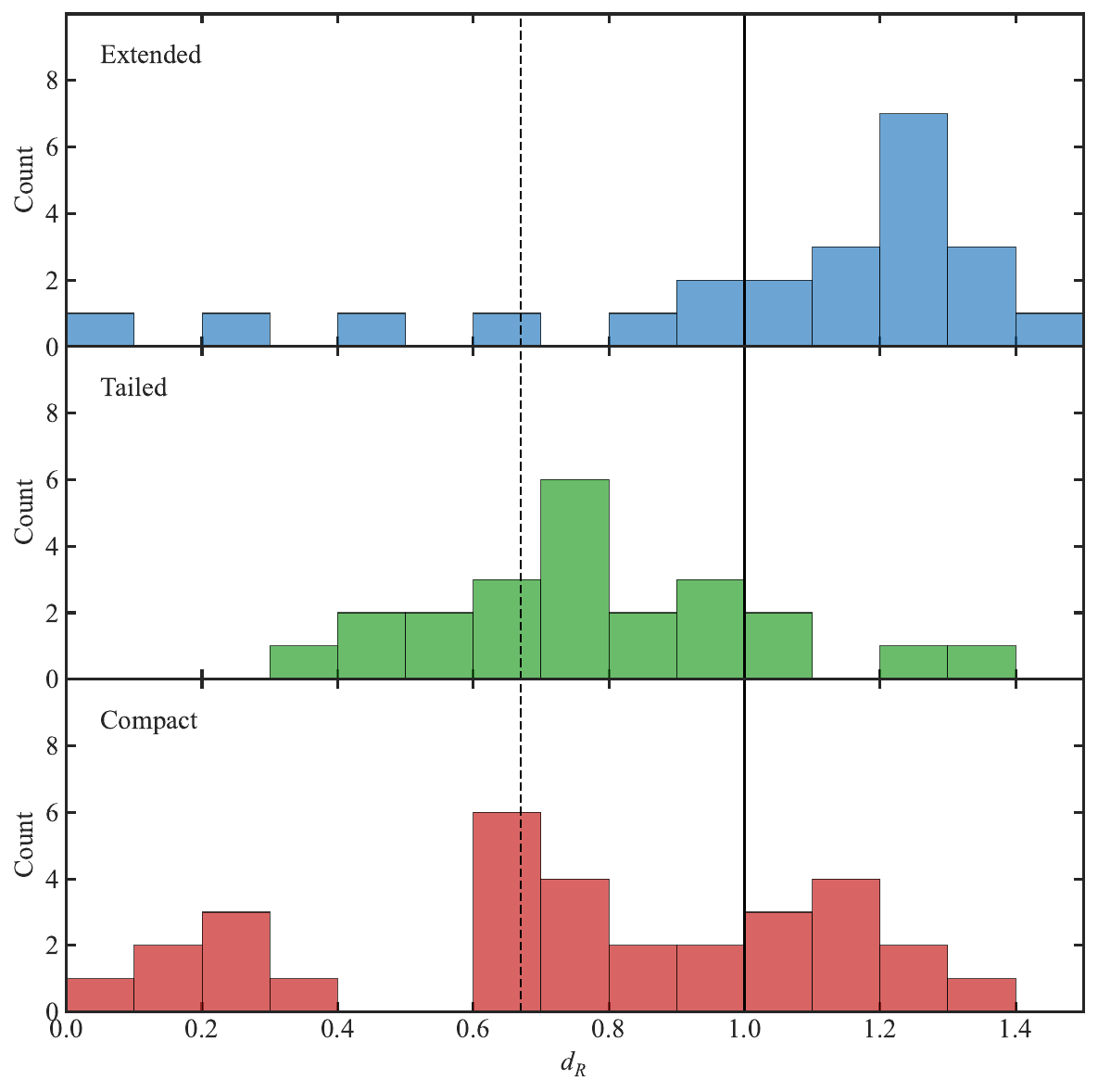}
\caption{$d_R$ distributions for the three radio morphology types: Extended (\textbf{top}), Tailed (\textbf{middle}), and~Compact (\textbf{bottom}). In~each panel, the~vertical black dashed line denotes the location of $R_{500}$, while the vertical black solid line marks $R_{200}$.}
\label{fig:cet}
\end{figure}

From the $d_R$ distribution histograms of these sources (Figure \ref{fig:cet}), we find that the majority (18 out of 25, $\sim$ 72\%) of extended (E) radio sources are located beyond $d_R > 1$. These sources reside in the cluster outskirts, and~their extended morphologies are consequently less likely to be disturbed by the dense cluster environment. In~contrast, most (18 out of 23, $\sim$ 78\%) of the tailed (T) radio sources are concentrated in the intermediate region ($0.4 < d_R < 1.0$). This distribution strongly suggests that the intense interaction between radio-emitting activity and the dense environment, which results in the characteristic swept-back morphology, primarily occurs in the cluster's main infall region. Furthermore, the~compact (C) radio sources exhibit a spatial distribution peaking around $R \approx R_{500}$ ($d_R \approx 0.65$). This indicates that the dense cluster environment not only affects the morphology of radio sources but may also play a role in modulating $\mathrm{AGN}$ activity.
\vspace{-3pt}

\begin{figure}[H]
\includegraphics[width=0.8\textwidth]{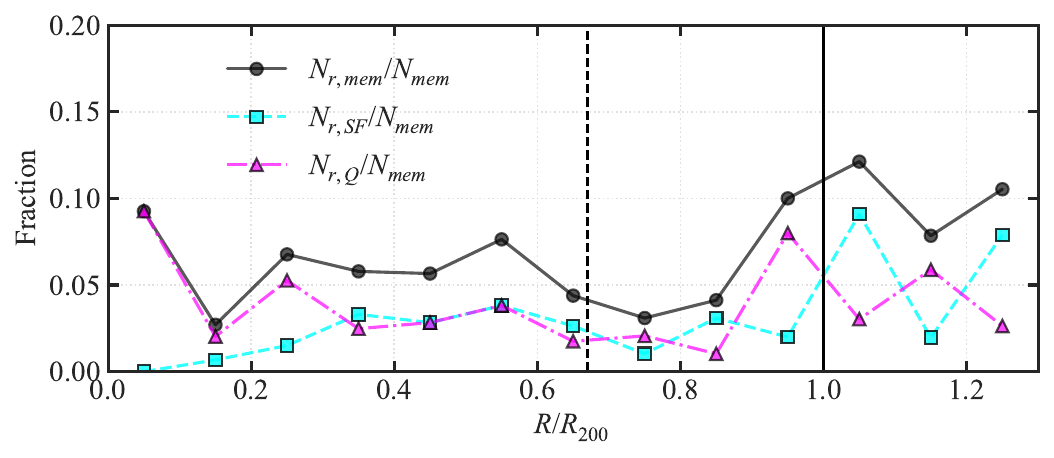} \\
\includegraphics[width=0.8\textwidth]{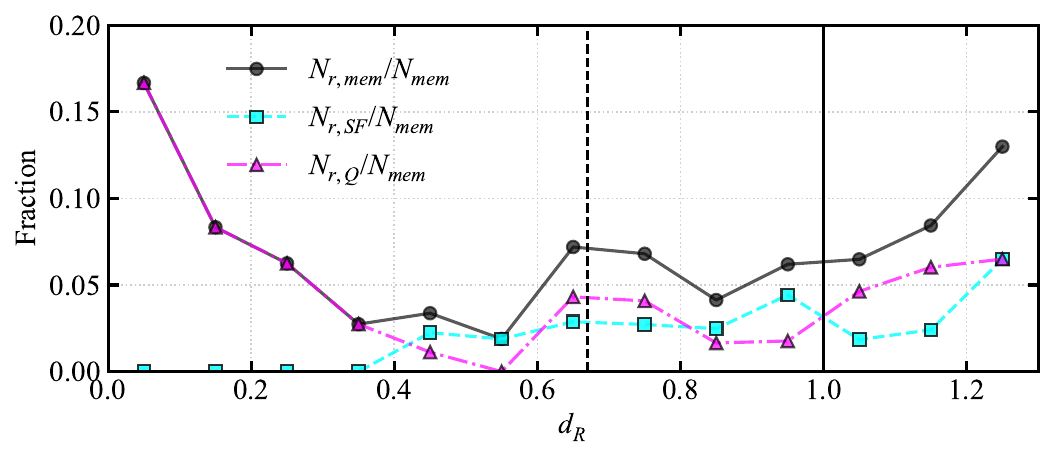}
\caption{(\textbf{Upper panel}): Fraction of radio sources as a function of $R/R_{200}$ for various populations in the Coma cluster. Black circles represent the fraction of all radio sources relative to the total optical member population. Cyan squares and magenta triangles denote the fractions of star-forming and quiescent galaxies with radio emission, respectively, both normalized to the total optical population.
Vertical black dashed and solid lines indicate the positions of $R_{500}$ and $R_{200}$.
(\textbf{Bottom panel}): Radio source fractions as a function of the $d_R$. Symbol types, colors and lines are consistent with the upper~panel.}
\label{fig:hist}
\end{figure}

We next examined the proportion of radio sources across different cluster environments. When using the projected radial distance ($R/R_{200}$) alone (the upper panel of Figure~\ref{fig:hist}), the~fraction of radio activity appears similar in $\mathrm{SF}$ and quiescent ($\mathrm{Q}$) member galaxies, except~in the very central region where $\mathrm{SF}$ galaxies are rare. {This interpretation is further supported by HI observations of Coma galaxies. \citet{Healy2021} found that HI deficiency increases significantly within $R \approx 0.5 R_{200}$, while galaxies at $0.6 < R/R_{200} < 1.0$ generally remain HI-normal. Our finding of an enhanced radio fraction near $R_{500}$ ($\sim0.65 R_{200}$) corresponds precisely to this transition zone. It suggests that as galaxies penetrate the $R_{500}$ boundary, the~rapid removal of HI reservoirs (as documented by \cite{Healy2021}) is accompanied by a transient boost in radio emission likely driven by ram-pressure-induced gas compression before the galaxies ultimately become HI-deficient and quenched in the cluster core.}
However, a~complicating factor is the observation that the overall radio fraction shows relatively high proportions both around $R_{200}$ and within $R_{500}$ on this plot, a~trend that is difficult to interpret due to projection~effects.

\vspace{-3pt}

By adopting the phase-space distance $d_R$ as the $\mathrm{x}$-axis (the lower panel of {Figure \ref{fig:cet}}), we effectively mitigate projection effects to a certain extent, and~the environmental trends become clearer. The~radio fraction exhibits relatively high proportions at three distinct locations in phase space: the core region ($d_R \approx 0$), an~intermediate peak around $d_R \approx 0.65$ (corresponding to $R_{500}$), and~the infall region beyond the virial radius ($d_R > 1$). At~the cluster center, the~enhanced radio fraction is predominantly driven by massive quiescent galaxies, implying more frequent accretion activity in the supermassive black holes within these~systems.

\textls[-15]{The peak in radio activity observed around the intermediate distance ($R_{500}$, or~$d_R \approx 0.65$)} is predominantly caused by compact and tailed radio sources (as detailed in Figure~\ref{fig:hist}). This peak coincides with the region where the density of the $\mathrm{ICM}$ increases rapidly, thereby maximizing the ram pressure experienced by infalling galaxies. This is strongly supported by the concentration of tailed sources here. Moreover, $\mathrm{AGN}$ accretion activity may be transiently boosted by external perturbations, possibly related to the recent merging event involving the infalling $\mathrm{NGC~4839}$ group \citep{Lyskova2019, Churazov2021}.

Conversely, the~generally declining trend in the radio fraction towards the cluster center (for $d_R < 1$) and the reduced activity beyond $R_{200}$ (where $d_R$ is large) are also understandable. The~dense cluster environment ultimately strips away the cold gas necessary for both efficient star formation and sustained $\mathrm{AGN}$ accretion, leading to the long-term suppression of radio~activity.

\section{Conclusions} \label{sec:conclusions}

In this study, we presented a systematic analysis of the radio activity within the Coma cluster by combining deep $144\,\mathrm{MHz}$ observations from LOFAR (LoTSS DR2) with optical spectroscopy and photometry from DESI and SDSS. We identified 79 cluster members with radio emission within $1.35 R_{200}$ (1.8 deg$^2$) and investigated their physical properties and environmental~dependence.

Our main conclusions are summarized as follows:

{\noindent\textbf{{(1) The Number and Morphology of Radio Sources. 
}
Thanks to LOFAR's superior sensitivity, we present the most comprehensive catalog of radio members of the Coma Cluster to date.
Our sample (79 members) more than doubles the population identified at 1.4 GHz by \citet{Miller2009} and substantially expands upon the 24 members recently surveyed with uGMRT by \citet{Lal2022}.
The observed source count distribution (log$N$-log$S$) shows a slight excess above the average trend{, but} it is not contributed by radio activity within the cluster.
Our 79 radio members are classified into 31 compact sources, 25 extended sources, and~23 tailed structures. Compared with existing reports \citep{2025ApJ...983..138R}, the~proportion of extended sources is significantly higher, while the proportion of compact sources is markedly lower, regardless of whether the system is merging or relaxing. However, the~proportion of tailed sources is comparable to that of the merging system.}

\noindent\textbf{(2) Bimodal Nature of Radio Hosts.} The radio population exhibits a clear bimodality in the star formation rate versus stellar mass plane, separating into massive quiescent galaxies and active star-forming systems. This optical state is strongly correlated with radio morphology. Compact (C) sources are predominantly hosted by massive, quiescent galaxies on the red sequence. Their radio emission exceeds the level expected from star formation alone, confirming it is dominated by central $\mathrm{AGN}$ activity. Tailed (T) sources are largely associated with star-forming galaxies. Their alignment with the star-forming main sequence and their specific morphologies indicate that their radio emission originates primarily from star formation activity occurring within gas stripped by ram~pressure.

\noindent\textbf{(3) Phase-Space Evolution.} Using the phase-space distance ($d_R$) as a proxy for infall time, we mapped the distinct environmental histories of these populations. Extended (E) sources are preferentially located in the cluster outskirts ($d_R > 1$), where they are entering the cluster environment for the first time and remain relatively undisturbed. Tailed (T) sources are concentrated in the infall region ($0.4 < d_R < 1.0$), confirming that the transformation of galaxy morphology via ram pressure stripping is most active during the first infall into the high-density $\mathrm{ICM}$.

\noindent\textbf{(4) Environmental Modulation of Activity.} The fractions of {radio-bright galaxies are} not uniform across the cluster. We observed elevated radio fractions in the core (due to massive $\mathrm{AGN}$) and, notably, a~secondary peak around $R_{500}$ ($d_R \approx 0.65$). This intermediate peak, populated by both compact and tailed sources, suggests that the dense environment at $R_{500}$ acts as a ``trigger zone.'' Here, increasing ram pressure drives morphological disturbances (tails), while external perturbations (potentially linked to substructure mergers like NGC 4839) may transiently boost $\mathrm{AGN}$ accretion before the galaxies are ultimately quenched in the~core.

In summary, our multi-wavelength analysis demonstrates that the Coma cluster environment plays a dual role in galaxy evolution. It initially promotes distinct radio morphologies and enhances activity during infall (at $\sim R_{500}$), before~ultimately stripping the gas reservoirs required to sustain star formation and $\mathrm{AGN}$ activity in the cluster~center.

\vspace{6pt}




\authorcontributions{Conceptualization, X.H. and H.Y.; methodology, X.H. and H.Y.; software, X.H.; validation, X.H., H.Y. and T.P.; formal analysis, X.H.; investigation, X.H. and H.D.; resources, X.H. and T.P.; data curation, X.H.; writing---original draft preparation, X.H.; writing---review and editing, H.Y., H.Z. and E.M.; visualization, X.H.; supervision, H.Y., H.Z. and C.L.; project administration, H.Y.; funding acquisition, H.Y., H.Z. and C.L. All authors have read and agreed to the published version of the manuscript.}

\funding{This work has been supported by the National Natural Science Foundation of China No. 12573003, the~National Key Research and Development Program of China (No. 2023YFC2206704), the~China Manned Space Program with grant No. CMS-CSST-2025-A04 and the International Partnership Program of Chinese Academy of Sciences, Grant No. 013GJHZ2024015FN. H.Zou. acknowledges the support from the National Key R\&D Program of China (grant Nos. 2023YFA1607804 and 2025YFE0202300) and the National Natural Science Foundation of China (NSFC; grant Nos. 12120101003 and 12373010) and China Manned Space Project (No. CMS-CSST-2025-A06).}

\institutionalreview{Not applicable.}

\informedconsent{Not applicable.}

\dataavailability{The data used in this study are from publicly available surveys. The~LOFAR Two-Metre Sky Survey (LoTSS) DR2 data can be accessed via the LOFAR Surveys website at 
 \url{https://lofar-surveys.org/} (accessed on 28 May 2025). The~DESI Legacy Imaging Surveys (DR9) data are available at \url{https://www.legacysurvey.org/} (accessed on 28 November 2023). Spectroscopic data from the Dark Energy Spectroscopic Instrument (DESI) DR1 and the Sloan Digital Sky Survey (SDSS) are available through their respective archives at \url{https://data.desi.lbl.gov/}  (accessed on 29 June 2025) and \url{https://www.sdss.org/} (accessed on 29 June 2025).}

\acknowledgments{The authors thank the referees for their corrections and valuable suggestions to improve the~paper.}

\conflictsofinterest{The authors declare no conflicts of~interest.}

\appendixtitles{yes} 
\appendixstart
\appendix
\section[\appendixname~\thesection]{Physical Parameters of Host Galaxies of Radio Sources in the Coma Cluster}
{In this appendix, we present the properties of the 79 radio-detected galaxies within the Coma cluster. Table~\ref{tab:datatable} lists the coordinates, spectroscopic measurements from DESI, and~radio properties from LOFAR.}
\vspace{-3pt}

\setlength{\tabcolsep}{4pt}
\renewcommand{\arraystretch}{1.2}
\renewcommand{\aboverulesep}{.1pt}
\renewcommand{\belowrulesep}{.1pt}
\begin{table}[H]
\scriptsize
\caption{Physical Parameters of Host Galaxies of Radio Sources in the Coma Cluster 
}

\begin{adjustwidth}{-\extralength}{0cm}

\begin{tabularx}{\fulllength}{CCCCCCCCCCcc}

\toprule
\textbf{ID} & \textbf{RA} & \textbf{Dec} & \textbf{z} & \textbf{logSFR} & \textbf{logM$_*$} & \textbf{logL$_{144}$} & \textbf{g-r} & \textbf{SNR} & \textbf{A} & \textbf{M$_{type}$} & \textbf{Name}\\
\midrule

%

1 
 & 193.2233 & 28.3713 & 0.0236 & $-$1.0157 & 11.1675 & 23.2975 & 0.8106 & 159.4 & Q & E &  \\
2 & 193.2814 & 28.2412 & 0.0235 & $-$0.6578 & 8.6951 & 21.3763 & 0.2414 & 4.2 & SF & E &  \\
3 & 193.6383 & 27.6328 & 0.0244 & $-$1.5908 & 10.7169 & 21.8449 & 0.8852 & 10.0 & Q & E &  \\
4 & 193.7237 & 28.4170 & 0.0247 & $-$1.2597 & 10.5773 & 21.6050 & 0.7287 & 7.8 & Q & P &  \\
5 & 193.7535 & 27.4834 & 0.0232 & $-$0.5168 & 9.6409 & 20.9695 & 0.6976 & 6.6 & SF & T & \\
6 & 193.9491 & 28.2562 & 0.0271 & $-$2.7578 & 10.6619 & 21.7663 & 0.9521 & 15.9 & Q & P &  \\
7 & 194.0457 & 28.1632 & 0.0290 & $-$1.4242 & 10.5203 & 22.3104 & 0.7733 & 32.7 & Q & E &  \\
8 & 194.1160 & 26.9874 & 0.0215 & $-$2.0609 & 11.1503 & 21.3991 & 0.7584 & 7.1 & Q & P & NGC~4819 \\
9 & 194.1191 & 27.2913 & 0.0251 & 0.3423 & 10.0223 & 22.3951 & 0.5356 & 12.6 & SF & T & IC~3913 \\
10 & 194.2690 & 27.7730 & 0.0252 & $-$1.0280 & 10.0424 & 21.4552 & 0.7290 & 8.3 & Q & P &  \\
11 & 194.2975 & 29.0450 & 0.0251 & $-$2.6986 & 11.0690 & 21.3544 & 0.7843 & 13.1 & Q & P &  \\
12 & 194.3515 & 27.4978 & 0.0245 & $-$1.3934 & 11.4962 & 23.9337 & 0.8238 & 225.1 & Q & T & NGC~4839 \\
13 & 194.3832 & 28.4770 & 0.0226 & $-$1.8004 & 11.2456 & 21.3970 & 0.8209 & 7.7 & Q & P & NGC~4841 \\
14 & 194.3868 & 27.6103 & 0.0202 & $-$2.2230 & 10.8419 & 21.1927 & 0.8175 & 16.1 & Q & P & NGC~4840 \\
15 & 194.3911 & 28.4822 & 0.0209 & $-$2.4123 & 10.7735 & 20.8330 & 0.7794 & 5.1 & Q & P & NGC~4841B \\
16 & 194.3994 & 27.4932 & 0.0244 & $-$2.0729 & 10.8500 & 21.8726 & 0.8449 & 23.0 & Q & T & NGC~4842A \\
17 & 194.4861 & 27.0375 & 0.0247 & $-$2.2138 & 10.3507 & 21.2190 & 0.7963 & 6.8 & Q & P &  \\
18 & 194.4908 & 28.0615 & 0.0272 & $-$0.2679 & 10.0576 & 22.7905 & 0.5459 & 73.1 & SF & T & KUG 1255+283\\
19 & 194.5234 & 28.2425 & 0.0241 & 0.1863 & 10.8921 & 23.1299 & 0.7039 & 229.3 & SF & T & NGC~4848 \\
20 & 194.5295 & 26.7871 & 0.0240 & $-$0.6892 & 9.5043 & 21.4055 & 0.7250 & 4.8 & SF & P &  \\
21 & 194.5386 & 28.7086 & 0.0254 & $-$0.2710 & 10.5953 & 22.7259 & 0.6768 & 37.9 & SF & T & \\
22 & 194.5776 & 27.3108 & 0.0248 & 0.1650 & 9.8159 & 22.6033 & 0.5623 & 59.4 & SF & T &  2MASX J12581865+2718387\\
23 & 194.6064 & 28.1289 & 0.0274 & $-$0.1316 & 9.4701 & 21.5277 & 0.4726 & 12.5 & SF & P &  \\

24 & 194.6230 & 28.3012 & 0.0197 & $-$2.0608 & 9.7707 & 21.9931 & 0.6995 & 38.5 & Q & T & \\
25 & 194.6473 & 27.2647 & 0.0246 & 0.3400 & 9.5105 & 22.4461 & 0.3864 & 46.9 & SF & T & Mrk~56\\
26 & 194.6553 & 27.1765 & 0.0256 & 0.4868 & 9.7762 & 22.5896 & 0.4429 & 39.6 & SF & E & Mrk~57\\
27 & 194.6577 & 27.4639 & 0.0209 & $-$0.7475 & 9.8283 & 21.2547 & 0.7178 & 6.0 & SF & P &  \\

\bottomrule
\end{tabularx}
\end{adjustwidth}
\end{table}

\begin{table}[H]\ContinuedFloat
\scriptsize
\caption{\textit{Cont.}}
\begin{adjustwidth}{-\extralength}{0cm}

\begin{tabularx}{\fulllength}{CCCCCCCCCCcc}

\toprule
\textbf{ID} & \textbf{RA} & \textbf{Dec} & \textbf{z} & \textbf{logSFR} & \textbf{logM$_*$} & \textbf{logL$_{144}$} & \textbf{g-r} & \textbf{SNR} & \textbf{A} & \textbf{M$_{type}$} & \textbf{Name}\\
\midrule

28 & 194.6601 & 27.5441 & 0.0199 & $-$2.2441 & 9.9729 & 20.9931 & 0.7714 & 10.2 & Q & P &  \\
29 & 194.6661 & 26.7594 & 0.0249 & $-$0.2461 & 9.0988 & 22.2670 & 0.3091 & 9.4 & SF & E &  \\
30 & 194.6910 & 28.5430 & 0.0214 & $-$0.6492 & 8.0268 & 21.0120 & 0.2144 & 5.2 & SF & P &  \\
31 & 194.7333 & 27.8334 & 0.0250 & $-$0.6321 & 10.8093 & 22.3000 & 0.7057 & 34.9 & Q & T & IC~3949 \\
32 & 194.7576 & 28.2253 & 0.0268 & $-$2.3230 & 10.9372 & 21.2362 & 0.8268 & 10.3 & Q & P &  \\
33 & 194.7576 & 26.8157 & 0.0239 & $-$2.4211 & 11.1191 & 21.1945 & 0.8588 & 7.1 & Q & P & NGC~4859 \\
34 & 194.7721 & 27.6443 & 0.0181 & 0.4105 & 9.6307 & 22.1972 & 0.4258 & 69.3 & SF & T & Mrk~58\\
35 & 194.7831 & 27.8550 & 0.0220 & $-$2.1787 & 10.5967 & 21.2490 & 0.8464 & 6.7 & Q & P &  IC~3960 \\
36 & 194.8047 & 27.9770 & 0.0227 & $-$1.7094 & 10.8568 & 21.7360 & 0.8194 & 10.6 & Q & P & NGC~4864 \\
37 & 194.8195 & 27.1061 & 0.0280 & $-$0.3123 & 10.1353 & 22.3609 & 0.8254 & 58.6 & SF & E &  \\
38 & 194.8464 & 28.4886 & 0.0233 & $-$0.1814 & 9.5231 & 21.6367 & 0.4914 & 10.3 & SF & E &  \\
39 & 194.8725 & 27.8501 & 0.0228 & $-$2.3703 & 10.5905 & 21.4116 & 0.8449 & 8.0 & Q & P &  IC~3976 \\
40 & 194.9130 & 28.8955 & 0.0204 & $-$2.5466 & 10.9182 & 21.5458 & 0.8250 & 29.6 & Q & P & IC~3990 \\
41 & 194.9173 & 28.6308 & 0.0178 & $-$0.2191 & 9.3282 & 22.2849 & 0.2772 & 7.7 & SF & T & 2MASX J12594009+2837507\\
42 & 195.0185 & 27.9876 & 0.0213 & $-$3.0756 & 10.7331 & 20.8999 & 0.7918 & 5.2 & Q & P & NGC~4886 \\
43 & 195.0381 & 27.8664 & 0.0178 & $-$0.3794 & 9.5411 & 21.3453 & 0.5153 & 25.4 & SF & T & \\
44 & 195.0737 & 27.9553 & 0.0231 & $-$1.9421 & 10.8803 & 21.5432 & 0.8097 & 17.5 & Q & P & NGC~4898A \\
45 & 195.0747 & 28.2024 & 0.0284 & $-$1.3262 & 11.2895 & 21.5054 & 0.8103 & 14.5 & Q & P & NGC~4895 \\
46 & 195.1303 & 28.9505 & 0.0233 & $-$0.4629 & 9.1845 & 21.5757 & 0.5040 & 4.5 & SF & E &  \\
47 & 195.1398 & 27.5041 & 0.0186 & $-$0.3401 & 8.8462 & 21.5959 & 0.2983 & 11.1 & SF & T & \\
48 & 195.1404 & 27.6377 & 0.0250 & 0.4759 & 9.4902 & 22.5324 & 0.3495 & 51.8 & SF & T & KUG 1258+279A\\
49 & 195.1580 & 28.0575 & 0.0258 & $-$2.2341 & 10.5447 & 23.2354 & 0.5974 & 203.6 & Q & T &  IC~4040 \\
50 & 195.1649 & 29.0193 & 0.0243 & $-$1.4084 & 10.9222 & 22.4582 & 0.7034 & 5.4 & Q & E &  IC~842 \\
51 & 195.1782 & 27.9713 & 0.0213 & $-$2.1353 & 10.7590 & 21.6363 & 0.8231 & 12.3 & Q & P &  IC~4042 \\
52 & 195.2027 & 28.0907 & 0.0232 & $-$2.3876 & 10.8586 & 21.1329 & 0.8192 & 5.3 & Q & P &  IC~4045 \\
53 & 195.2335 & 27.7908 & 0.0266 & $-$1.9604 & 11.3670 & 23.4662 & 0.7630 & 72.3 & Q & E & NGC~4911 \\
54 & 195.3530 & 29.3097 & 0.0238 & $-$0.7136 & 11.0558 & 22.9374 & 0.7694 & 343.4 & Q & E &  \\
55 & 195.3589 & 27.8860 & 0.0183 & $-$0.5649 & 9.5893 & 22.0810 & 0.8207 & 8.1 & SF & T & NGC~4921 \\
56 & 195.3900 & 29.1306 & 0.0244 & $-$1.5341 & 11.1295 & 22.5343 & 0.8700 & 87.4 & Q & E &  IC~843 \\
57 & 195.4307 & 29.0446 & 0.0238 & $-$1.8624 & 11.0517 & 22.5015 & 0.7774 & 22.9 & Q & E &  IC~4088 \\
58 & 195.4899 & 28.0058 & 0.0258 & $-$2.4214 & 11.2327 & 22.5212 & 0.8588 & 75.4 & Q & T & NGC~4927 \\
59 & 195.5175 & 29.2534 & 0.0244 & $-$2.0235 & 10.9985 & 21.2754 & 0.8273 & 12.4 & Q & P &  \\
60 & 195.5328 & 27.6483 & 0.0231 & 0.6086 & 10.1078 & 22.4365 & 0.5984 & 43.9 & SF & T & NGC~4926A \\
61 & 195.5361 & 28.3871 & 0.0253 & $-$1.3320 & 10.7821 & 21.3179 & 0.8186 & 10.4 & Q & P &  \\
62 & 195.5534 & 28.2148 & 0.0273 & 0.4652 & 9.9039 & 22.5737 & 0.4530 & 38.7 & SF & T & \\
63 & 195.6069 & 28.8581 & 0.0224 & $-$0.2017 & 8.6330 & 21.3629 & 0.2562 & 9.0 & SF & P &  \\
64 & 195.8177 & 28.0303 & 0.0204 & $-$2.0619 & 10.7009 & 21.2667 & 0.7033 & 10.7 & Q & P & NGC~4934 \\
65 & 196.1489 & 28.6277 & 0.0224 & $-$0.0207 & 9.3745 & 21.5319 & 0.4092 & 12.6 & SF & E &  \\
66 & 196.4729 & 28.1124 & 0.0246 & $-$1.2821 & 10.3171 & 21.9081 & 0.7629 & 18.3 & Q & E &  \\
67 & 196.6517 & 27.8729 & 0.0209 & $-$0.1819 & 10.3712 & 22.6134 & 0.7172 & 99.0 & SF & E &  \\
68 & 195.0339 & 27.9769 & 0.0215 & $-$0.6327 & 11.4979 & 21.8058 & 0.9295 & 46.6 & Q & E & NGC~4889 \\
69 & 194.3800 & 26.5122 & 0.0241 & 0.0311 & 10.3631 & 22.5834 & 0.6695 & 57.3 & SF & E & IC~837  \\
70 & 194.6664 & 26.7595 & 0.0249 & $-$0.1239 & 8.9040 & 22.2666 & 0.1889 & 9.4 & SF & E &  \\
71 & 194.8988 & 27.9593 & 0.0239 & $-$1.0247 & 11.4851 & 24.1498 & 0.7986 & 707.3 & Q & E & NGC~4874 \\
72 & 193.4747 & 28.1867 & 0.0256 & $-$0.4953 & 10.0399 & 22.0574 & 0.6107 & 6.8 & SF & T & \\
73 & 193.7299 & 27.4127 & 0.0262 & $-$1.7811 & 11.3985 & 21.7893 & 0.8635 & 12.0 & Q & E & NGC~4798 \\

74 & 194.6801 & 28.9101 & 0.0278 & 0.0657 & 10.2042 & 21.4637 & 0.6192 & 5.1 & SF & P &  \\
75 & 195.1487 & 27.5742 & 0.0170 & -0.9065 & 10.1042 & 21.8469 & 0.7380 & 38.2 & Q & E &  \\
76 & 195.2148 & 28.0429 & 0.0292 & -1.8349 & 11.1536 & 21.6187 & 0.8274 & 16.0 & Q & P & NGC~4908 \\

77 & 196.0768 & 28.4681 & 0.0258 & -1.3036 & 10.2698 & 23.4086 & 0.7104 & 358.6 & Q & T & \\
78 & 196.0945 & 28.8108 & 0.0265 & 0.3497 & 9.7524 & 22.3147 & 0.3471 & 18.3 & SF & E &  \\
79 & 196.1106 & 27.3044 & 0.0184 & -0.0358 & 9.3695 & 21.9423 & 0.3349 & 10.6 & SF & E &  \\
\bottomrule
\end{tabularx}
\end{adjustwidth}
\label{tab:datatable}
\footnotesize{Notes 
. Column 
(1): Radio source ID. Column (2)--(3): Right Ascension and Declination (J2000) in decimal degrees. Column (4): Spectroscopic redshift from the DESI survey. Column (5): Logarithm of the star formation rate in units of $M_\odot\,\mathrm{yr}^{-1}$. Column (6): Logarithm of the stellar mass in units of $M_\odot$. Column (7): Logarithm of the radio luminosity at 144 MHz in units of $\mathrm{W\,Hz^{-1}}$. Column (8): Optical color $(g-r)$ corrected for Galactic extinction. Column (9): Peak signal-to-noise ratio (SNR) of the LOFAR radio emission. Column (10): Star-forming activity: star-forming (SF) or quiescent (Q). Column (11): Radio morphology: C (Compact), E (Extended), or~T (Tailed). Column (12): Galaxy name, when available.}
\end{table}

\section[\appendixname~\thesection]{Galaxies with Large Radio-Optical Offsets}
\label{a2}
{In this section, we present the multi-wavelength images for cluster members exhibiting a significant positional offset ($6^{\prime\prime}$) between their radio emission centroid and optical center. Such offsets are often indicative of environmental processing, where the relativistic plasma is displaced from the stellar disk due to the ram pressure of the intracluster medium (ICM).}

\vspace{-3pt}
\begin{figure}[H]

    \includegraphics[width=0.45\linewidth]{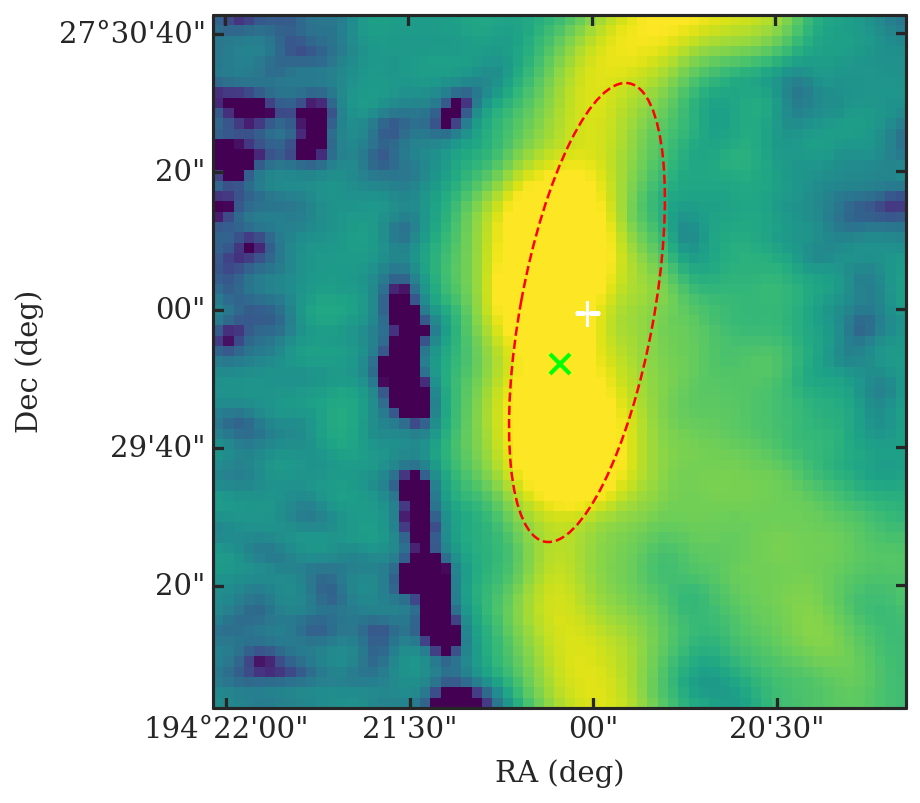}
    \includegraphics[width=0.45\linewidth]{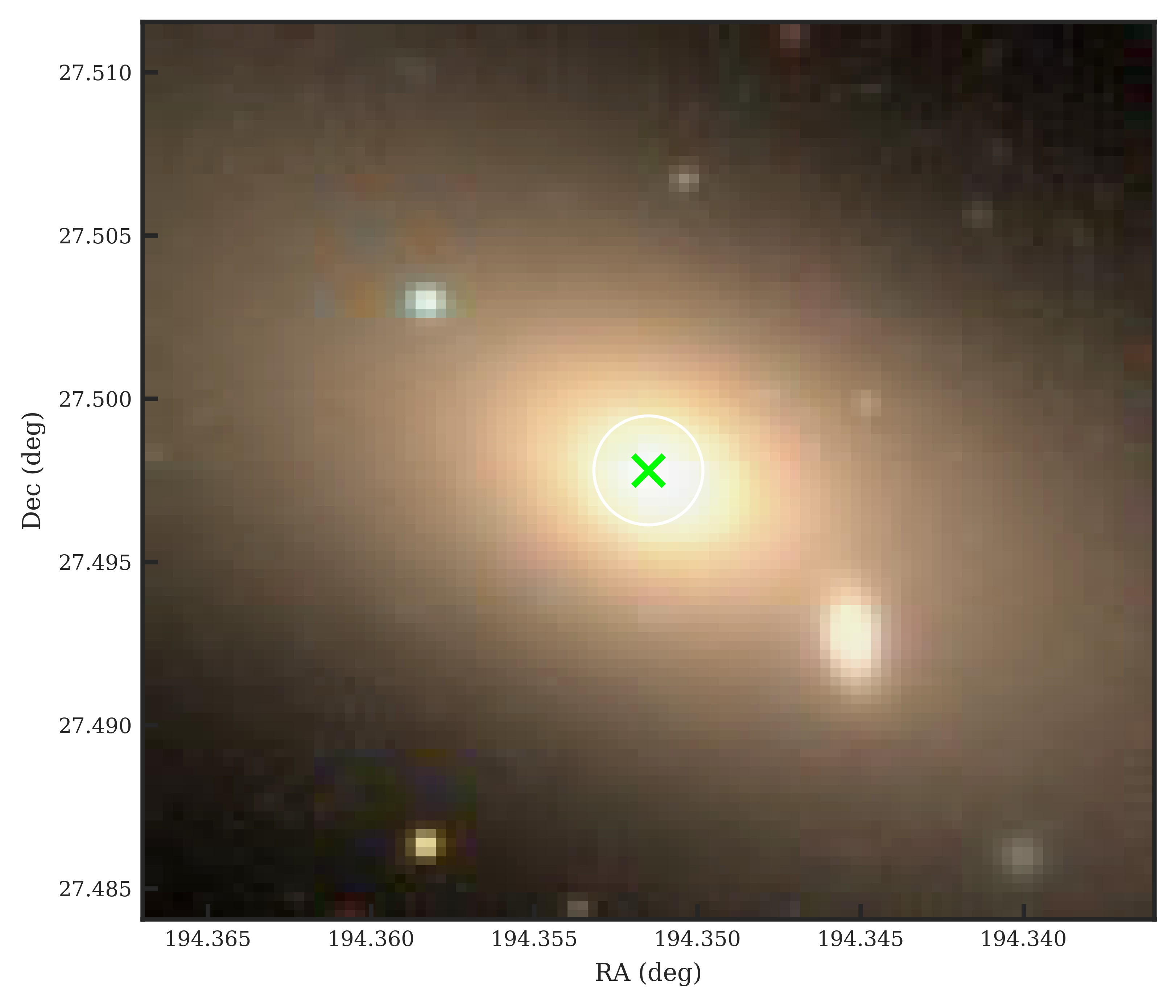}
    
    \includegraphics[width=0.45\linewidth]{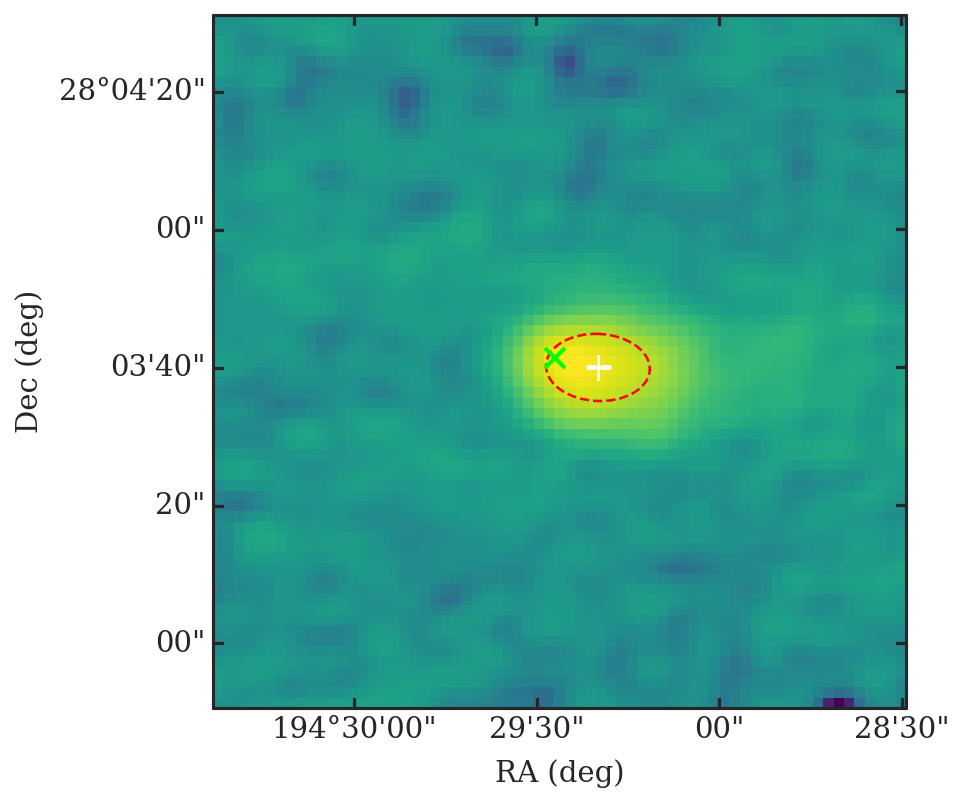}
    \includegraphics[width=0.45\linewidth]{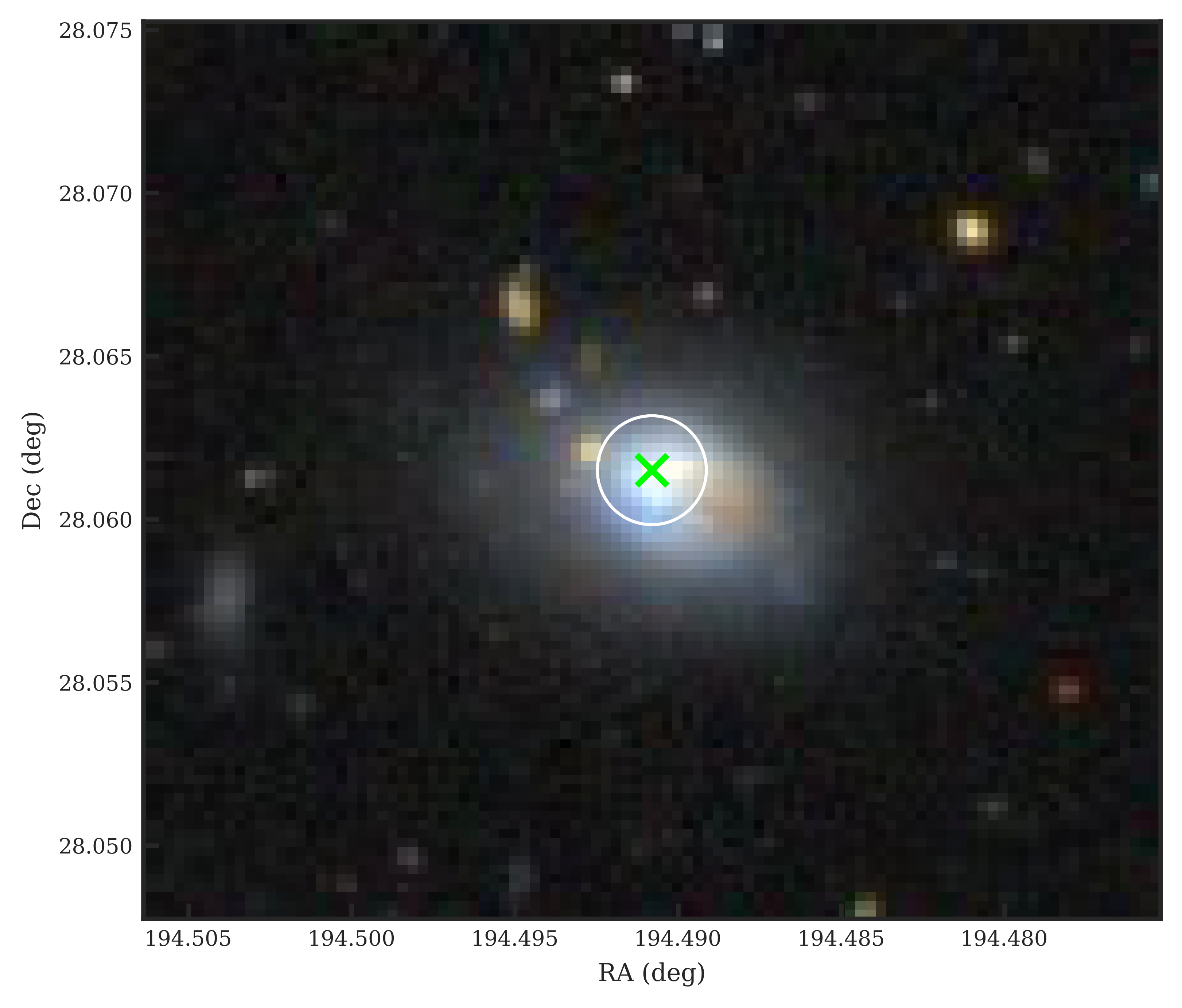}
    
    \includegraphics[width=0.45\linewidth]{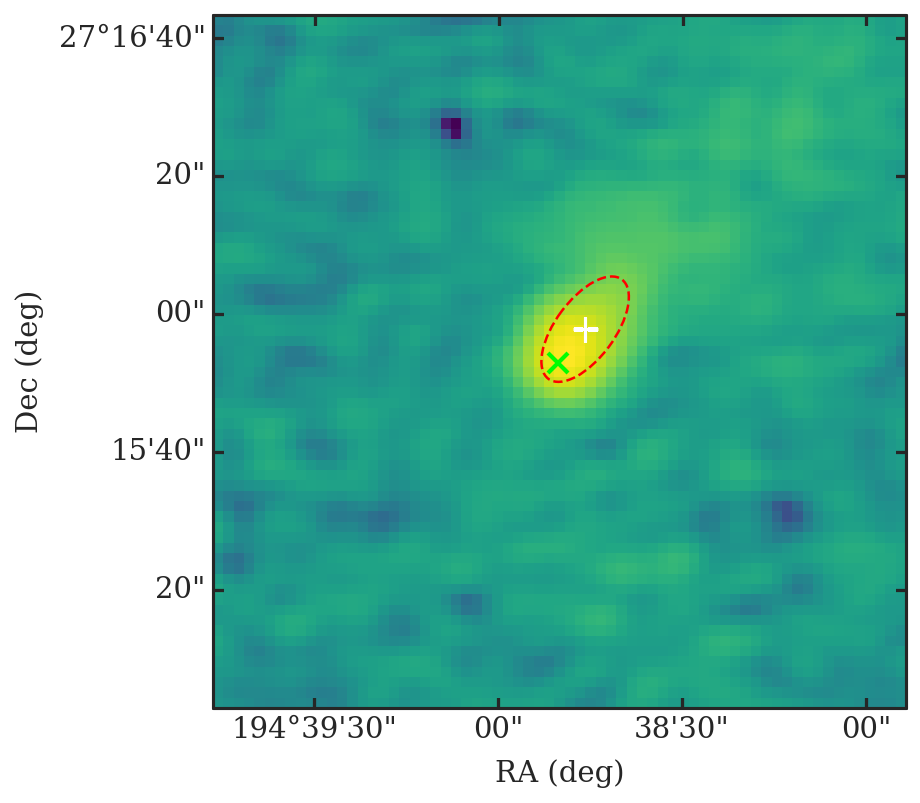}
    \includegraphics[width=0.45\linewidth]{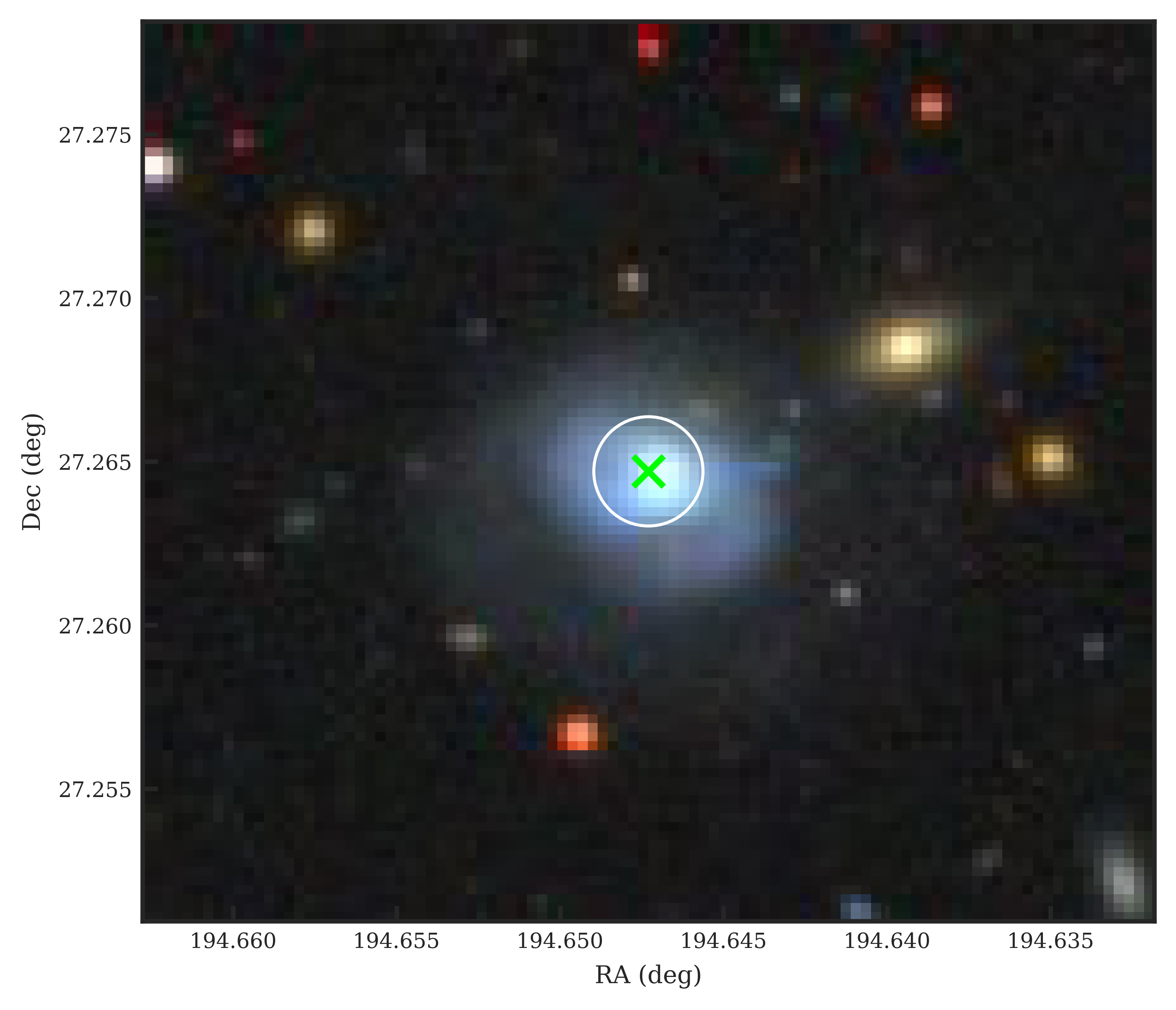}
    
    \includegraphics[width=0.45\linewidth]{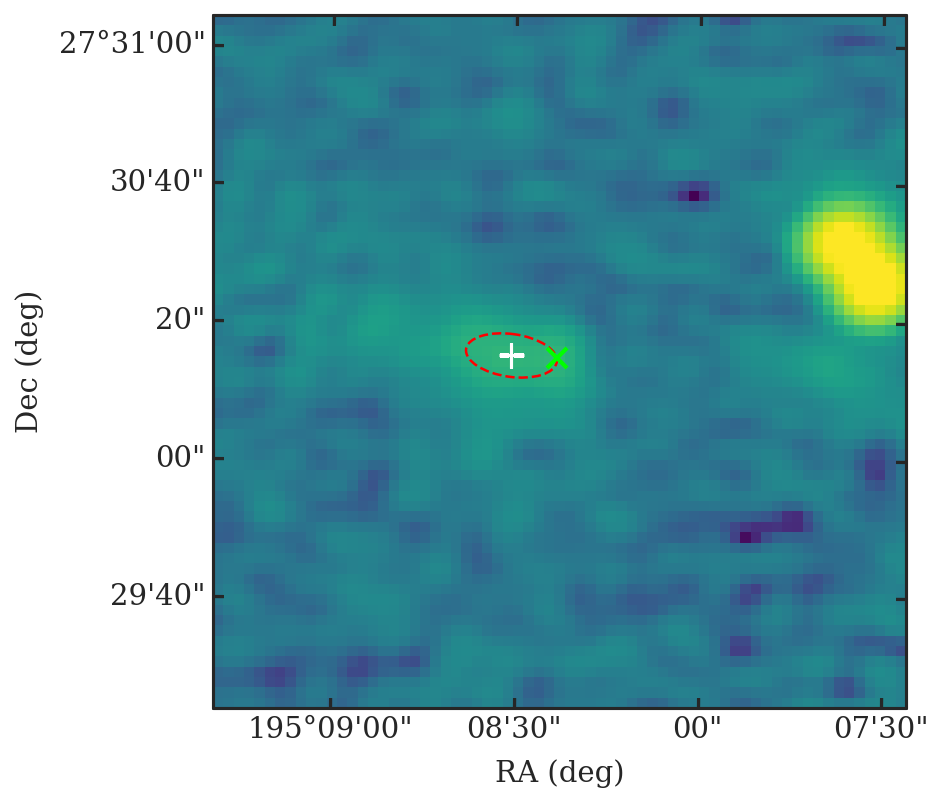}
    \includegraphics[width=0.45\linewidth]{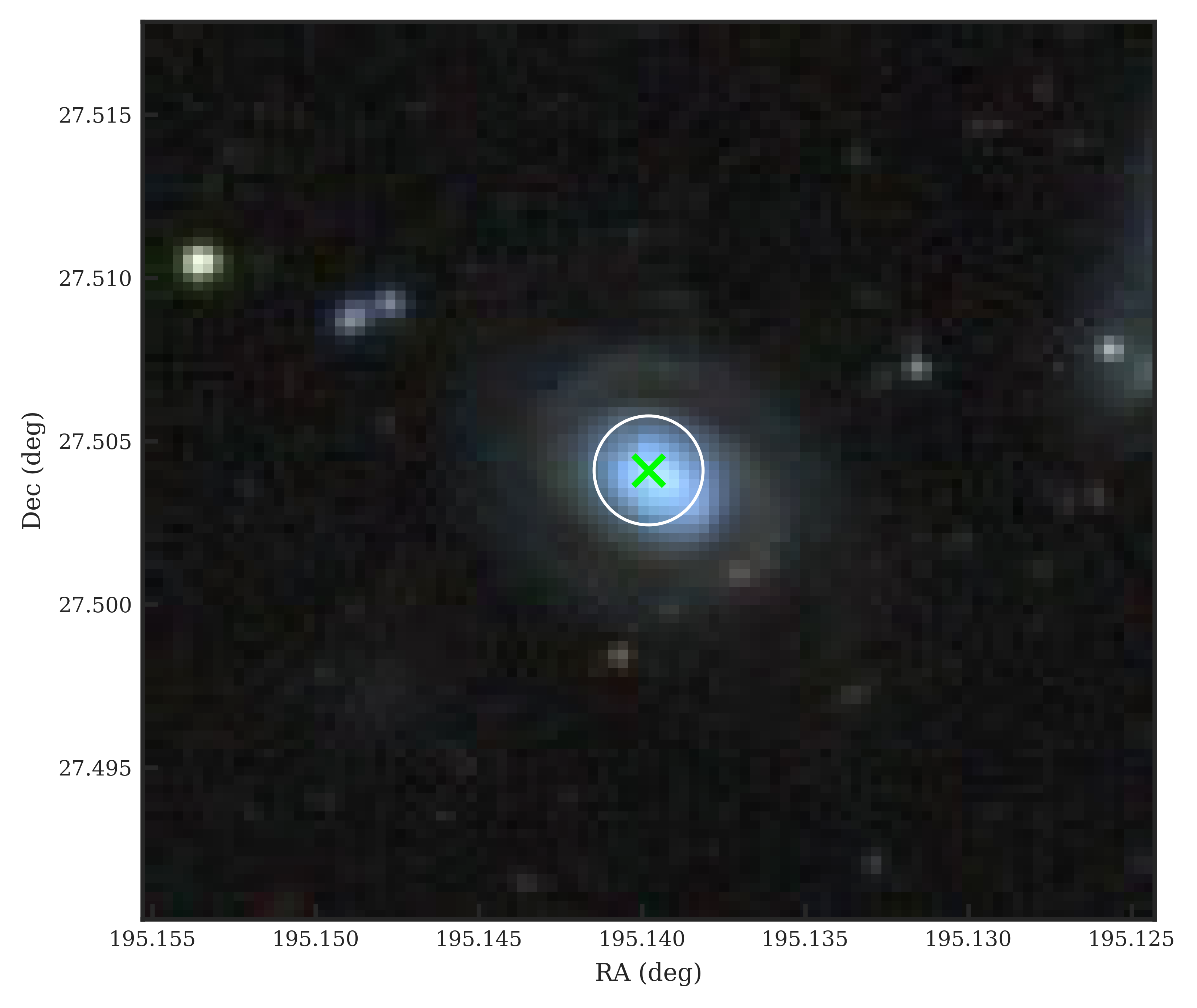}
  \caption{{Radio 
  and optical postage stamps for the sub-sample of Coma cluster members with a radio-optical offset exceeding $6^{\prime\prime}$. The~left panels display LOFAR 144 MHz radio continuum images (100$^{\prime\prime} \times 100^{\prime\prime}$), while the right panels show the corresponding optical images from DESI. In~all panels, green ``x'' symbols indicate the optical galactic centers, and~white crosses denote the centroids of the radio emission. The~red dashed ellipses in the radio images represent the morphology detections of the radio emission. The~white circles in the optical images have a radius of $6^{\prime\prime}$. The~sub-sample includes: ID 12 (offset = $8.41^{\prime\prime}$), ID 18 ($6.26^{\prime\prime}$), ID 25 ($6.24^{\prime\prime}$), and~ID 47 ($6.59^{\prime\prime}$).}}
  \label{fig:offsets}
\end{figure}
\unskip

\begin{figure}[H]
    \includegraphics[width=0.45\linewidth]{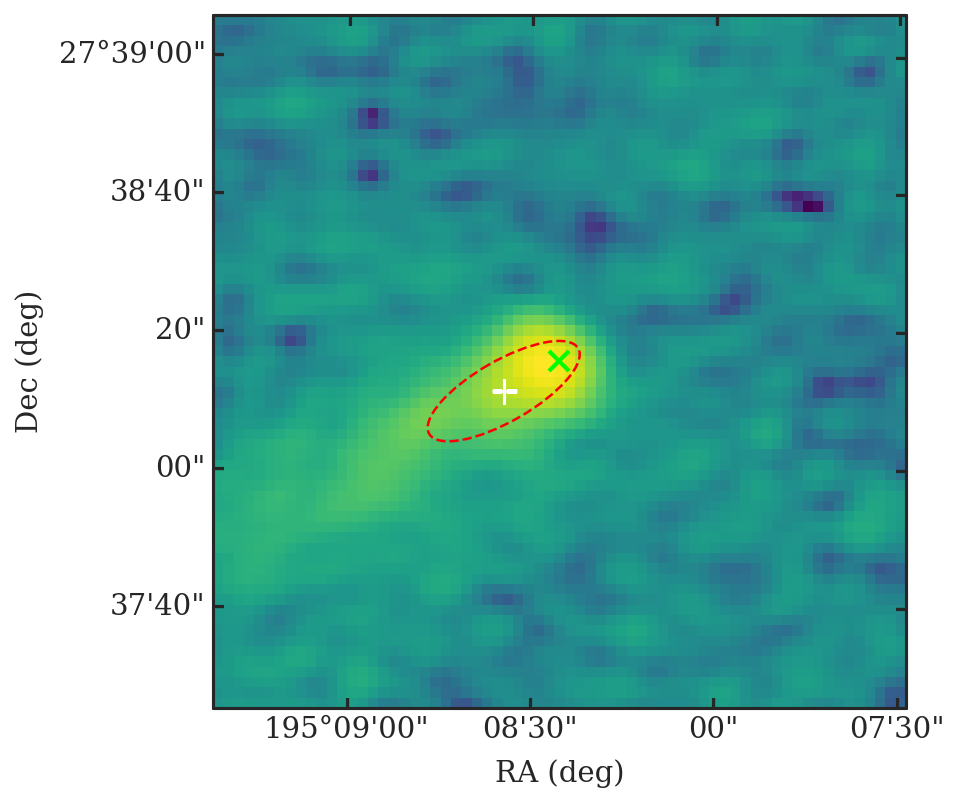}
    \includegraphics[width=0.45\linewidth]{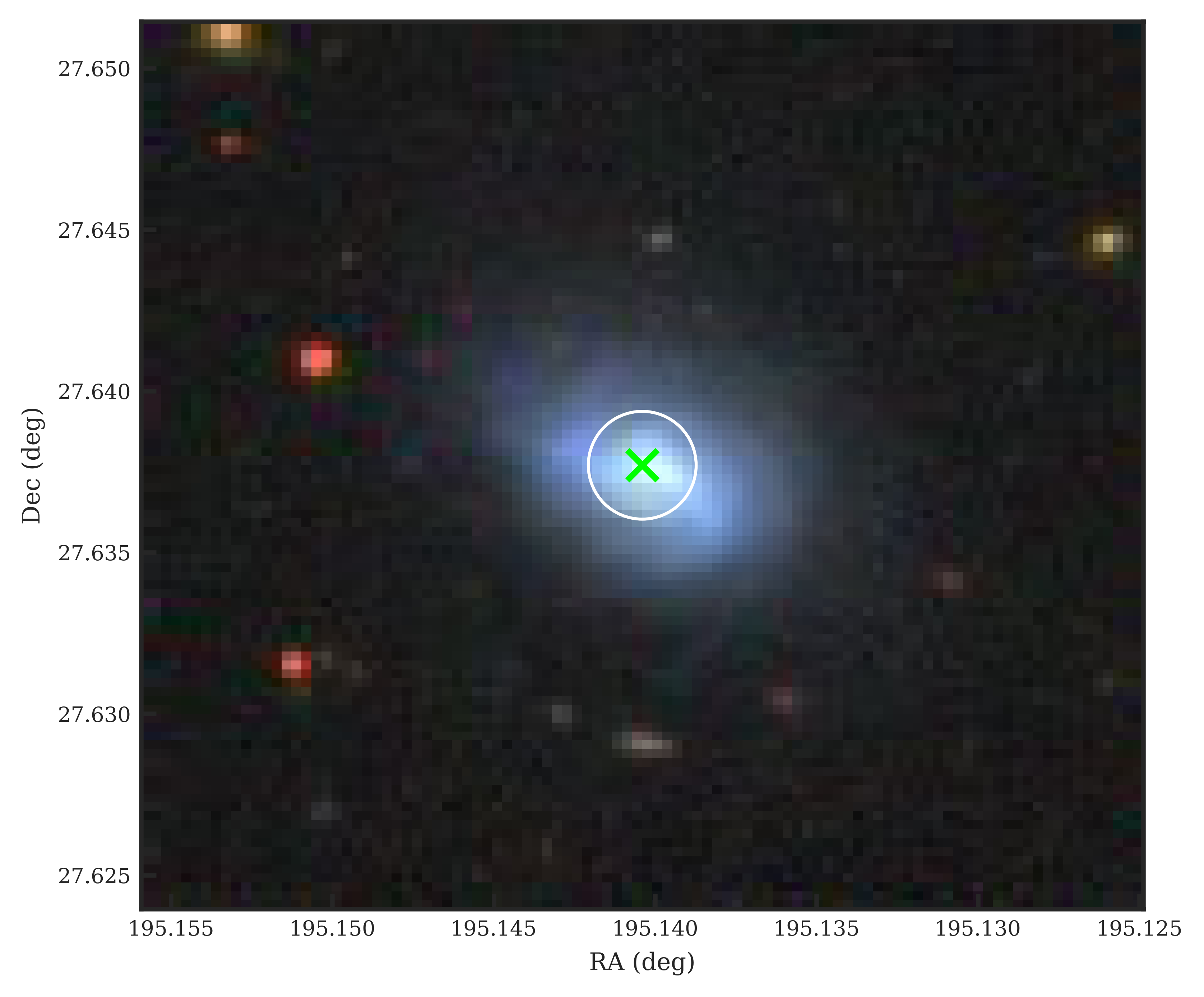}
    
    \includegraphics[width=0.45\linewidth]{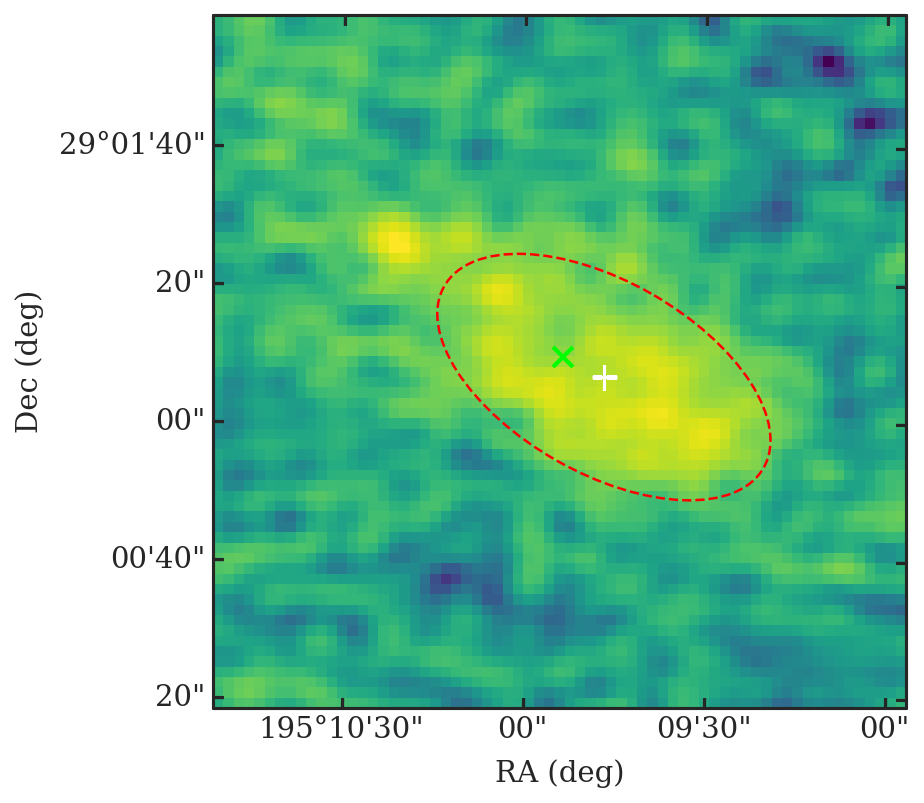}
    \includegraphics[width=0.45\linewidth]{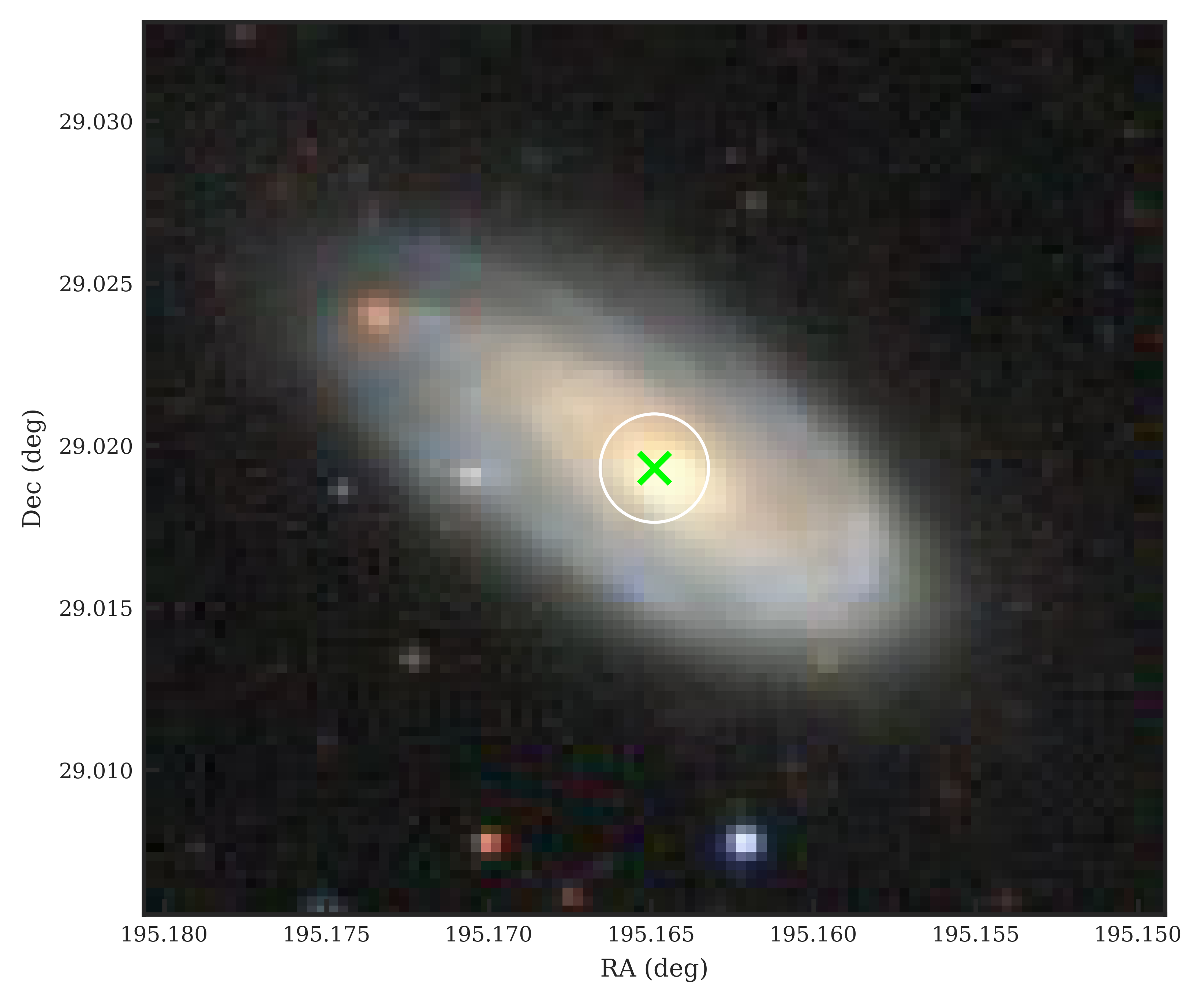}
    
    \includegraphics[width=0.45\linewidth]{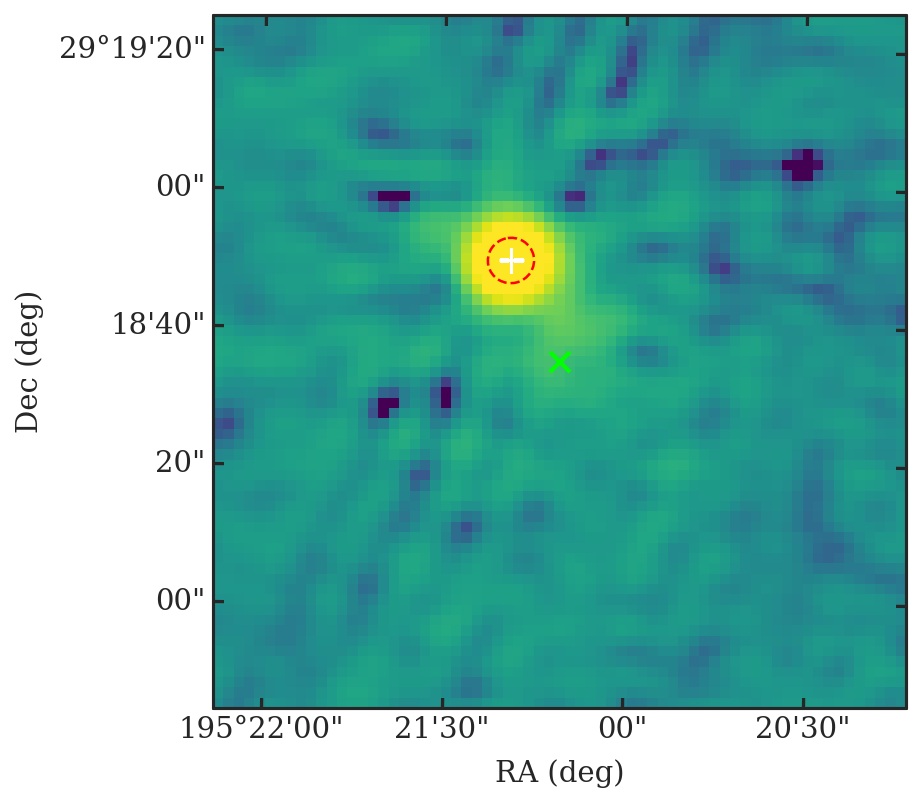}
    \includegraphics[width=0.45\linewidth]{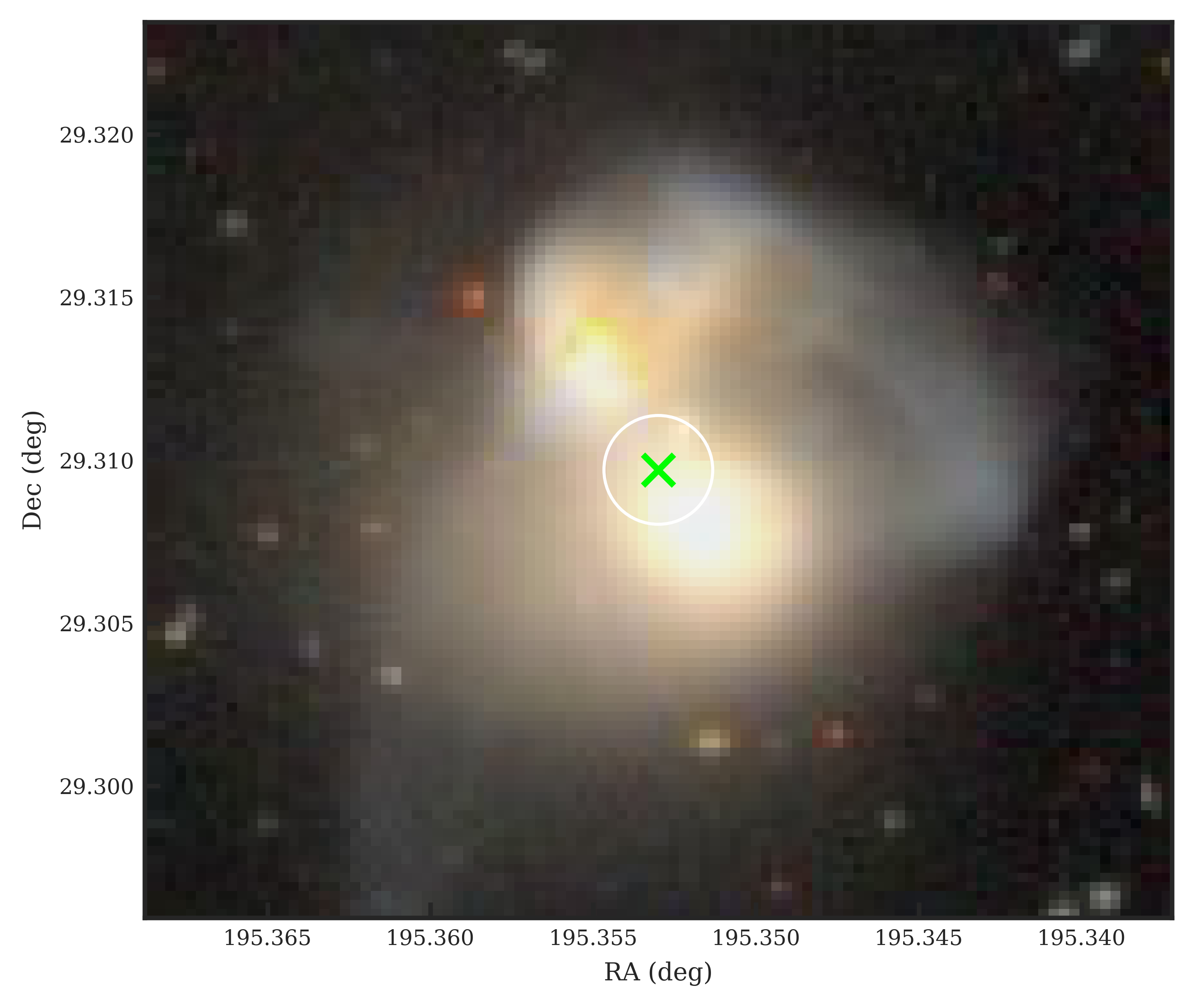}
  \caption{{Radio 
and optical postage stamps for the sub-sample of Coma cluster members with a radio-optical offset exceeding $6^{\prime\prime}$ (Continued). The~sub-sample includes: ID 48 (offset = $9.29^{\prime\prime}$), ID 50($6.55^{\prime\prime}$), and~ID 54 ($16.28^{\prime\prime}$).}}
  \label{fig:offsets2}
\end{figure}

\begin{adjustwidth}{-\extralength}{0cm}

\reftitle{References}



\PublishersNote{}
\end{adjustwidth}

\begin{thebibliography}{999}

\bibitem[{Gunn} and {Gott}(1972)]{1972ApJ...176....1G}
{Gunn}, J.E.; {Gott}, J.R. III.
\newblock {On the Infall of Matter Into Clusters of Galaxies and Some Effects
  on Their Evolution}.
\newblock {\em {Astrophys. J.}} {\bf 1972}, {\em 176},~1.
\newblock {\url{https://doi.org/10.1086/151605}}.

\bibitem[{Boselli} and {Gavazzi}(2006)]{2006PASP..118..517B}
{Boselli}, A.; {Gavazzi}, G.
\newblock {Environmental Effects on Late-Type Galaxies in Nearby Clusters}.
\newblock {\em {Publ. Astron. Soc. Pac.}} {\bf 2006}, {\em 118},~517--559.
\newblock {\url{https://doi.org/10.1086/500691}}.

\bibitem[{Bregman}(1990)]{Bregman1990}
{Bregman}, J.N.
\newblock {Continuum radiation from active galactic nuclei}.
\newblock {\em {Astron. Astrophys. Rev.}} {\bf 1990}, {\em 2},~125--166.
\newblock {\url{https://doi.org/10.1007/BF00872765}}.

\bibitem[{Turner} and {Ho}(1994)]{Turner1994}
{Turner}, J.L.; {Ho}, P.T.P.
\newblock {Bright Radio Continuum Emission from Star Formation in the Cores of
  Nearby Spiral Galaxies}.
\newblock {\em {Astrophys. J.}} {\bf 1994}, {\em 421},~122.
\newblock {\url{https://doi.org/10.1086/173631}}.

\bibitem[{Klein} et~al.(2018){Klein}, {Lisenfeld}, and {Verley}]{Klein2018}
{Klein}, U.; {Lisenfeld}, U.; {Verley}, S.
\newblock {Radio synchrotron spectra of star-forming galaxies}.
\newblock {\em {Astron. Astrophys.}} {\bf 2018}, {\em 611},~A55.
\newblock {\url{https://doi.org/10.1051/0004-6361/201731673}}.

\bibitem[{Panessa} et~al.(2019){Panessa}, {Baldi}, {Laor}, {Padovani}, {Behar},
  and {McHardy}]{Panessa2019}
{Panessa}, F.; {Baldi}, R.D.; {Laor}, A.; {Padovani}, P.; {Behar}, E.;
  {McHardy}, I.
\newblock {The origin of radio emission from radio-quiet active galactic
  nuclei}.
\newblock {\em Nat. Astron.} {\bf 2019}, {\em 3},~387--396.
\newblock {\url{https://doi.org/10.1038/s41550-019-0765-4}}.

\bibitem[{Rickel} et~al.(2025){Rickel}, {Moravec}, {Gordon}, {Hardcastle},
  {Pierce}, {Bilton}, and {Roberts}]{2025ApJ...983..138R}
{Rickel}, M.; {Moravec}, E.; {Gordon}, Y.A.; {Hardcastle}, M.J.; {Pierce},
  J.C.S.; {Bilton}, L.E.; {Roberts}, I.D.
\newblock {The Merging Galaxy Cluster Environment Affects the Morphology of
  Radio Active Galactic Nuclei}.
\newblock {\em {Astrophys. J.}} {\bf 2025}, {\em 983},~138.
\newblock {\url{https://doi.org/10.3847/1538-4357/adbb5e}}.

\bibitem[{Ledlow} and {Owen}(1996)]{Ledlow1996}
{Ledlow}, M.J.; {Owen}, F.N.
\newblock {20 CM VLA Survey of Abell Clusters of Galaxies. VI. Radio/Optical
  Luminosity Functions}.
\newblock {\em Astron. J.} {\bf 1996}, {\em 112},~9.
\newblock {\url{https://doi.org/10.1086/117985}}.

\bibitem[{Wing} and {Blanton}(2011)]{Wing2011}
{Wing}, J.D.; {Blanton}, E.L.
\newblock {Galaxy Cluster Environments of Radio Sources}.
\newblock {\em Astron. J.} {\bf 2011}, {\em 141},~88.
\newblock {\url{https://doi.org/10.1088/0004-6256/141/3/88}}.

\bibitem[{van Weeren} et~al.(2019){van Weeren}, {de Gasperin}, {Akamatsu},
  {Br{\"u}ggen}, {Feretti}, {Kang}, {Stroe}, and {Zandanel}]{Weeren2019}
{van Weeren}, R.J.; {de Gasperin}, F.; {Akamatsu}, H.; {Br{\"u}ggen}, M.;
  {Feretti}, L.; {Kang}, H.; {Stroe}, A.; {Zandanel}, F.
\newblock {Diffuse Radio Emission from Galaxy Clusters}.
\newblock {\em Space Sci. Rev.} {\bf 2019}, {\em 215},~16.
\newblock {\url{https://doi.org/10.1007/s11214-019-0584-z}}.

\bibitem[{Salim} and {Rich}(2010)]{Salim2010}
{Salim}, S.; {Rich}, R.M.
\newblock {Star Formation Signatures in Optically Quiescent Early-type
  Galaxies}.
\newblock {\em Astrophys. J. Lett.} {\bf 2010}, {\em
  714},~L290--L294.
\newblock {\url{https://doi.org/10.1088/2041-8205/714/2/L290}}.

\bibitem[{Capetti} et~al.(2022){Capetti}, {Brienza}, {Balmaverde}, {Best},
  {Baldi}, {Drabent}, {G{\"u}rkan}, {Rottgering}, {Tasse}, and
  {Webster}]{Capetti2022}
{Capetti}, A.; {Brienza}, M.; {Balmaverde}, B.; {Best}, P.N.; {Baldi}, R.D.;
  {Drabent}, A.; {G{\"u}rkan}, G.; {Rottgering}, H.J.A.; {Tasse}, C.;
  {Webster}, B.
\newblock {The LOFAR view of giant, early-type galaxies: Radio emission from
  active nuclei and star formation}.
\newblock {\em Astron. Astrophys.} {\bf 2022}, {\em 660},~A93.
\newblock {\url{https://doi.org/10.1051/0004-6361/202142911}}.

\bibitem[{Best}(2004)]{Best2004}
{Best}, P.N.
\newblock {The environmental dependence of radio-loud AGN activity and star
  formation in the 2dFGRS}.
\newblock {\em {Mon. Not. R. Astron. Soc.}} {\bf 2004}, {\em 351},~70--82.
\newblock {\url{https://doi.org/10.1111/j.1365-2966.2004.07752.x}}.

\bibitem[{Comerford} et~al.(2020){Comerford}, {Negus},
  {M{\"u}ller-S{\'a}nchez}, {Eracleous}, {Wylezalek}, {Storchi-Bergmann},
  {Greene}, {Barrows}, {Nevin}, {Roy}, and {Stemo}]{Comerford2020}
{Comerford}, J.M.; {Negus}, J.; {M{\"u}ller-S{\'a}nchez}, F.; {Eracleous}, M.;
  {Wylezalek}, D.; {Storchi-Bergmann}, T.; {Greene}, J.E.; {Barrows}, R.S.;
  {Nevin}, R.; {Roy}, N.;  et~al.
\newblock {A Catalog of 406 AGNs in MaNGA: A Connection between Radio-mode AGNs
  and Star Formation Quenching}.
\newblock {\em {Astrophys. J.}} {\bf 2020}, {\em 901},~159.
\newblock {\url{https://doi.org/10.3847/1538-4357/abb2ae}}.

\bibitem[{Churazov} et~al.(2023){Churazov}, {Khabibullin}, {Bykov}, {Lyskova},
  and {Sunyaev}]{2023A&A...670A.156C}
\textls[-15]{{Churazov}, E.; {Khabibullin}, I.; {Bykov}, A.M.; {Lyskova}, N.; {Sunyaev}, R.
\newblock {Tempestuous life beyond R$_{500}$: X-ray view on the Coma cluster
  with SRG/eROSITA. II. Shock and relic}.
\newblock {\em {Astron. Astrophys.}} {\bf 2023}, {\em 670},~A156.
\newblock {\url{https://doi.org/10.1051/0004-6361/202244021}}.}

\bibitem[{Dey} et~al.(2019){Dey}, {Schlegel}, {Lang}, {Blum}, {Burleigh},
  {Fan}, {Findlay}, {Finkbeiner}, {Herrera}, {Juneau}, {Landriau}, {Levi},
  {McGreer}, {Meisner}, {Myers}, {Moustakas}, {Nugent}, {Patej}, {Schlafly},
  {Walker}, {Valdes}, {Weaver}, {Y{\`e}che}, {Zou}, {Zhou}, {Abareshi},
  {Abbott}, {Abolfathi}, {Aguilera}, {Alam}, {Allen}, {Alvarez}, {Annis},
  {Ansarinejad}, {Aubert}, {Beechert}, {Bell}, {BenZvi}, {Beutler}, {Bielby},
  {Bolton}, {Brice{\~n}o}, {Buckley-Geer}, {Butler}, {Calamida}, {Carlberg},
  {Carter}, {Casas}, {Castander}, {Choi}, {Comparat}, {Cukanovaite}, {Delubac},
  {DeVries}, {Dey}, {Dhungana}, {Dickinson}, {Ding}, {Donaldson}, {Duan},
  {Duckworth}, {Eftekharzadeh}, {Eisenstein}, {Etourneau}, {Fagrelius},
  {Farihi}, {Fitzpatrick}, {Font-Ribera}, {Fulmer}, {G{\"a}nsicke},
  {Gaztanaga}, {George}, {Gerdes}, {Gontcho}, {Gorgoni}, {Green}, {Guy},
  {Harmer}, {Hernandez}, {Honscheid}, {Huang}, {James}, {Jannuzi}, {Jiang},
  {Joyce}, {Karcher}, {Karkar}, {Kehoe}, {Kneib}, {Kueter-Young}, {Lan},
  {Lauer}, {Le Guillou}, {Le Van Suu}, {Lee}, {Lesser}, {Perreault Levasseur},
  {Li}, {Mann}, {Marshall}, {Mart{\'\i}nez-V{\'a}zquez}, {Martini}, {du Mas des
  Bourboux}, {McManus}, {Meier}, {M{\'e}nard}, {Metcalfe},
  {Mu{\~n}oz-Guti{\'e}rrez}, {Najita}, {Napier}, {Narayan}, {Newman}, {Nie},
  {Nord}, {Norman}, {Olsen}, {Paat}, {Palanque-Delabrouille}, {Peng},
  {Poppett}, {Poremba}, {Prakash}, {Rabinowitz}, {Raichoor}, {Rezaie},
  {Robertson}, {Roe}, {Ross}, {Ross}, {Rudnick}, {Safonova}, {Saha},
  {S{\'a}nchez}, {Savary}, {Schweiker}, {Scott}, {Seo}, {Shan}, {Silva},
  {Slepian}, {Soto}, {Sprayberry}, {Staten}, {Stillman}, {Stupak}, {Summers},
  {Sien Tie}, {Tirado}, {Vargas-Maga{\~n}a}, {Vivas}, {Wechsler}, {Williams},
  {Yang}, {Yang}, {Yapici}, {Zaritsky}, {Zenteno}, {Zhang}, {Zhang}, {Zhou},
  and {Zhou}]{2019AJ....157..168D}
{Dey}, A.; {Schlegel}, D.J.; {Lang}, D.; {Blum}, R.; {Burleigh}, K.; {Fan}, X.;
  {Findlay}, J.R.; {Finkbeiner}, D.; {Herrera}, D.; {Juneau}, S.;  et~al.
\newblock {Overview of the DESI Legacy Imaging Surveys}.
\newblock {\em {Astron. J.}} {\bf 2019}, {\em 157},~168.
\newblock {\url{https://doi.org/10.3847/1538-3881/ab089d}}.

\bibitem[{DESI Collaboration} et~al.(2025){DESI Collaboration}, {Abdul-Karim},
  {Adame}, {Aguado}, {Aguilar}, {Ahlen}, {Alam}, {Aldering}, {Alexander},
  {Alfarsy}, {Allen}, {Allende Prieto}, {Alves}, {Anand}, {Andrade},
  {Armengaud}, {Avila}, {Aviles}, {Awan}, {Bailey}, {Baleato Lizancos},
  {Ballester}, {Bault}, {Bautista}, {BenZvi}, {Beraldo e Silva},
  {Bermejo-Climent}, {Beutler}, {Bianchi}, {Blake}, {Blum}, {Bolton}, {Bonici},
  {Brieden}, {Brodzeller}, {Brooks}, {Buckley-Geer}, {Burtin}, {Canning},
  {Carnero Rosell}, {Carr}, {Carrilho}, {Casas}, {Castander}, {Cereskaite},
  {Cervantes-Cota}, {Chaussidon}, {Chaves-Montero}, {Chen}, and
  {Chen}]{2025arXiv250314745D}
{DESI Collaboration}.; {Abdul-Karim}, M.; {Adame}, A.G.; {Aguado}, D.;
  {Aguilar}, J.; {Ahlen}, S.; {Alam}, S.; {Aldering}, G.; {Alexander}, D.M.;
  {Alfarsy}, R.;  et~al.
\newblock {Data Release 1 of the Dark Energy Spectroscopic Instrument}.
\newblock {\em arXiv e-prints} {\bf 2025}, p. arXiv:2503.14745.
\newblock {\url{https://doi.org/10.48550/arXiv.2503.14745}}.

\bibitem[{Churazov} et~al.(2021){Churazov}, {Khabibullin}, {Lyskova},
  {Sunyaev}, and {Bykov}]{Churazov2021}
{Churazov}, E.; {Khabibullin}, I.; {Lyskova}, N.; {Sunyaev}, R.; {Bykov}, A.M.
\newblock {Tempestuous life beyond R$_{500}$: X-ray view on the Coma cluster
  with SRG/eROSITA. I. X-ray morphology, recent merger, and radio halo
  connection}.
\newblock {\em {Astron. Astrophys.}} {\bf 2021}, {\em 651},~A41.
\newblock {\url{https://doi.org/10.1051/0004-6361/202040197}}.

\bibitem[{Miller} et~al.(2009){Miller}, {Hornschemeier}, and
  {Mobasher}]{Miller2009}
{Miller}, N.A.; {Hornschemeier}, A.E.; {Mobasher}, B.
\newblock {A Deep Very Large Array Radio Continuum Survey of the Core and
  Outskirts of the Coma Cluster}.
\newblock {\em {Astron. J.}} {\bf 2009}, {\em 137},~4436--4449.
\newblock {\url{https://doi.org/10.1088/0004-6256/137/5/4436}}.

\bibitem[{Lal} et~al.(2022){Lal}, {Lyskova}, {Zhang}, {Venturi}, {Forman},
  {Jones}, {Churazov}, {van Weeren}, {Bonafede}, {Miller}, {Roberts}, {Bykov},
  {Di Mascolo}, {Br{\"u}ggen}, and {Brunetti}]{Lal2022}
{Lal}, D.V.; {Lyskova}, N.; {Zhang}, C.; {Venturi}, T.; {Forman}, W.R.;
  {Jones}, C.; {Churazov}, E.M.; {van Weeren}, R.J.; {Bonafede}, A.; {Miller},
  N.A.;  et~al.
\newblock {High-resolution, High-sensitivity, Low-frequency uGMRT View of Coma
  Cluster of Galaxies}.
\newblock {\em Astrophys. J.} {\bf 2022}, {\em 934},~170.
\newblock {\url{https://doi.org/10.3847/1538-4357/ac7a9b}}.

\bibitem[{Shimwell} et~al.(2017){Shimwell}, {R{\"o}ttgering}, {Best},
  {Williams}, {Dijkema}, {de Gasperin}, {Hardcastle}, {Heald}, {Hoang},
  {Horneffer}, {Intema}, {Mahony}, {Mandal}, {Mechev}, {Morabito}, {Oonk},
  {Rafferty}, {Retana-Montenegro}, {Sabater}, {Tasse}, {van Weeren},
  {Br{\"u}ggen}, {Brunetti}, {Chy{\.z}y}, {Conway}, {Haverkorn}, {Jackson},
  {Jarvis}, {McKean}, {Miley}, {Morganti}, {White}, {Wise}, {van Bemmel},
  {Beck}, {Brienza}, {Bonafede}, {Calistro Rivera}, {Cassano}, {Clarke},
  {Cseh}, {Deller}, {Drabent}, {van Driel}, {Engels}, {Falcke}, {Ferrari},
  {Fr{\"o}hlich}, {Garrett}, {Harwood}, {Heesen}, {Hoeft}, {Horellou},
  {Israel}, {Kapi{\'n}ska}, {Kunert-Bajraszewska}, {McKay}, {Mohan},
  {Orr{\'u}}, {Pizzo}, {Prandoni}, {Schwarz}, {Shulevski}, {Sipior}, {Smith},
  {Sridhar}, {Steinmetz}, {Stroe}, {Varenius}, {van der Werf}, {Zensus}, and
  {Zwart}]{Shimwell2017}
{Shimwell}, T.W.; {R{\"o}ttgering}, H.J.A.; {Best}, P.N.; {Williams}, W.L.;
  {Dijkema}, T.J.; {de Gasperin}, F.; {Hardcastle}, M.J.; {Heald}, G.H.;
  {Hoang}, D.N.; {Horneffer}, A.;  et~al.
\newblock {The LOFAR Two-metre Sky Survey. I. Survey description and
  preliminary data release}.
\newblock {\em {Astron. Astrophys.}} {\bf 2017}, {\em 598},~A104.
\newblock {\url{https://doi.org/10.1051/0004-6361/201629313}}.

\bibitem[{Shimwell} et~al.(2022){Shimwell}, {Hardcastle}, {Tasse}, {Best},
  {R{\"o}ttgering}, {Williams}, {Botteon}, {Drabent}, {Mechev}, {Shulevski},
  {van Weeren}, {Bester}, {Br{\"u}ggen}, {Brunetti}, {Callingham}, {Chy{\.z}y},
  {Conway}, {Dijkema}, {Duncan}, {de Gasperin}, {Hale}, {Haverkorn}, {Hugo},
  {Jackson}, {Mevius}, {Miley}, {Morabito}, {Morganti}, {Offringa}, {Oonk},
  {Rafferty}, {Sabater}, {Smith}, {Schwarz}, {Smirnov}, {O'Sullivan},
  {Vedantham}, {White}, {Albert}, {Alegre}, {Asabere}, {Bacon}, {Bonafede},
  {Bonnassieux}, {Brienza}, {Bilicki}, {Bonato}, {Calistro Rivera}, {Cassano},
  {Cochrane}, {Croston}, {Cuciti}, {Dallacasa}, {Danezi}, {Dettmar}, {Di
  Gennaro}, {Edler}, {En{\ss}lin}, {Emig}, {Franzen}, {Garc{\'\i}a-Vergara},
  {Grange}, {G{\"u}rkan}, {Hajduk}, {Heald}, {Heesen}, {Hoang}, {Hoeft},
  {Horellou}, {Iacobelli}, {Jamrozy}, {Jeli{\'c}}, {Kondapally}, {Kukreti},
  {Kunert-Bajraszewska}, {Magliocchetti}, {Mahatma}, {Ma{\l}ek}, {Mandal},
  {Massaro}, {Meyer-Zhao}, {Mingo}, {Mostert}, {Nair}, {Nakoneczny},
  {Nikiel-Wroczy{\'n}ski}, {Orr{\'u}}, {Pajdosz-{\'S}mierciak}, {Pasini},
  {Prandoni}, {van Piggelen}, {Rajpurohit}, {Retana-Montenegro}, {Riseley},
  {Rowlinson}, {Saxena}, {Schrijvers}, {Sweijen}, {Siewert}, {Timmerman},
  {Vaccari}, {Vink}, {West}, {Wo{\l}owska}, {Zhang}, and
  {Zheng}]{2022A&A...659A...1S}
{Shimwell}, T.W.; {Hardcastle}, M.J.; {Tasse}, C.; {Best}, P.N.;
  {R{\"o}ttgering}, H.J.A.; {Williams}, W.L.; {Botteon}, A.; {Drabent}, A.;
  {Mechev}, A.; {Shulevski}, A.;  et~al.
\newblock {The LOFAR Two-metre Sky Survey. V. Second data release}.
\newblock {\em {Astron. Astrophys.}} {\bf 2022}, {\em 659},~A1.
\newblock {\url{https://doi.org/10.1051/0004-6361/202142484}}.

\bibitem[{Almeida} et~al.(2023){Almeida}, {Anderson}, {Argudo-Fern{\'a}ndez},
  {Badenes}, {Barger}, {Barrera-Ballesteros}, {Bender}, {Benitez}, {Besser},
  {Bird}, {Bizyaev}, {Blanton}, {Bochanski}, {Bovy}, {Brandt}, {Brownstein},
  {Buchner}, {Bulbul}, {Burchett}, {Cano D{\'\i}az}, {Carlberg}, {Casey},
  {Chandra}, {Cherinka}, {Chiappini}, {Coker}, {Comparat}, {Conroy},
  {Contardo}, {Cortes}, {Covey}, {Crane}, {Cunha}, {Dabbieri}, {Davidson},
  {Davis}, {de Andrade Queiroz}, {De Lee}, {M{\'e}ndez Delgado}, {Demasi}, {Di
  Mille}, {Donor}, {Dow}, {Dwelly}, {Eracleous}, {Eriksen}, {Fan}, {Farr},
  {Frederick}, {Fries}, {Frinchaboy}, {G{\"a}nsicke}, {Ge}, {Gonz{\'a}lez
  {\'A}vila}, {Grabowski}, {Grier}, {Guiglion}, {Gupta}, {Hall}, {Hawkins},
  {Hayes}, {Hermes}, {Hern{\'a}ndez-Garc{\'\i}a}, {Hogg}, {Holtzman},
  {Ibarra-Medel}, {Ji}, {Jofre}, {Johnson}, {Jones}, {Kinemuchi}, {Kluge},
  {Koekemoer}, {Kollmeier}, {Kounkel}, {Krishnarao}, {Krumpe}, {Lacerna},
  {Lago}, {Laporte}, {Liu}, {Liu}, {Liu}, {Lopes}, {Macktoobian}, {Majewski},
  {Malanushenko}, {Maoz}, {Masseron}, {Masters}, {Matijevic}, {McBride},
  {Medan}, {Merloni}, {Morrison}, {Myers}, {M{\'e}sz{\'a}ros}, {Negrete},
  {Nidever}, {Nitschelm}, {Oravetz}, {Oravetz}, {Pan}, {Peng}, {Pinsonneault},
  {Pogge}, {Qiu}, {Ramirez}, {Rix}, {Fern{\'a}ndez Rosso}, {Runnoe}, {Salvato},
  {Sanchez}, {Santana}, {Saydjari}, {Sayres}, {Schlaufman}, {Schneider},
  {Schwope}, {Serna}, {Shen}, {Sobeck}, {Song}, {Souto}, {Spoo}, {Stassun},
  {Steinmetz}, {Straumit}, {Stringfellow}, {S{\'a}nchez-Gallego},
  {Taghizadeh-Popp}, {Tayar}, {Thakar}, {Tissera}, {Tkachenko}, {Hernandez
  Toledo}, {Trakhtenbrot}, {Fern{\'a}ndez-Trincado}, {Troup}, {Trump},
  {Tuttle}, {Ulloa}, {Vazquez-Mata}, {Vera Alfaro}, {Villanova}, {Wachter},
  {Weijmans}, {Wheeler}, {Wilson}, {Wojno}, {Wolf}, {Xue}, {Ybarra}, {Zari},
  and {Zasowski}]{2023ApJS..267...44A}
{Almeida}, A.; {Anderson}, S.F.; {Argudo-Fern{\'a}ndez}, M.; {Badenes}, C.;
  {Barger}, K.; {Barrera-Ballesteros}, J.K.; {Bender}, C.F.; {Benitez}, E.;
  {Besser}, F.; {Bird}, J.C.;  et~al.
\newblock {The Eighteenth Data Release of the Sloan Digital Sky Surveys:
  Targeting and First Spectra from SDSS-V}.
\newblock {\em {Astrophys. J. Suppl. Ser.}} {\bf 2023}, {\em 267},~44.
\newblock {\url{https://doi.org/10.3847/1538-4365/acda98}}.

\bibitem[{Hahn} et~al.(2023){Hahn}, {Wilson}, {Ruiz-Macias}, {Cole},
  {Weinberg}, {Moustakas}, {Kremin}, {Tinker}, {Smith}, {Wechsler}, {Ahlen},
  {Alam}, {Bailey}, {Brooks}, {Cooper}, {Davis}, {Dawson}, {Dey}, {Dey},
  {Eftekharzadeh}, {Eisenstein}, {Fanning}, {Forero-Romero}, {Frenk},
  {Gazta{\~n}aga}, {A Gontcho}, {Guy}, {Honscheid}, {Ishak}, {Juneau}, {Kehoe},
  {Kisner}, {Lan}, {Landriau}, {Le Guillou}, {Levi}, {Magneville}, {Martini},
  {Meisner}, {Myers}, {Nie}, {Norberg}, {Palanque-Delabrouille}, {Percival},
  {Poppett}, {Prada}, {Raichoor}, {Ross}, {Gaines}, {Saulder}, {Schlafly},
  {Schlegel}, {Sierra-Porta}, {Tarle}, {Weaver}, {Y{\`e}che}, {Zarrouk},
  {Zhou}, {Zhou}, and {Zou}]{Hahn2023}
{Hahn}, C.; {Wilson}, M.J.; {Ruiz-Macias}, O.; {Cole}, S.; {Weinberg}, D.H.;
  {Moustakas}, J.; {Kremin}, A.; {Tinker}, J.L.; {Smith}, A.; {Wechsler}, R.H.;
   et~al.
\newblock {The DESI Bright Galaxy Survey: Final Target Selection, Design, and
  Validation}.
\newblock {\em Astron. J.} {\bf 2023}, {\em 165},~253.
\newblock {\url{https://doi.org/10.3847/1538-3881/accff8}}.

\bibitem[{Serra} et~al.(2011){Serra}, {Diaferio}, {Murante}, and
  {Borgani}]{Serra2011}
{Serra}, A.L.; {Diaferio}, A.; {Murante}, G.; {Borgani}, S.
\newblock {Measuring the escape velocity and mass profiles of galaxy clusters
  beyond their virial radius}.
\newblock {\em {Mon. Not. R. Astron. Soc.}} {\bf 2011}, {\em 412},~800--816.
\newblock {\url{https://doi.org/10.1111/j.1365-2966.2010.17946.x}}.

\bibitem[{Serra} and {Diaferio}(2013)]{Serra2013}
{Serra}, A.L.; {Diaferio}, A.
\newblock {Identification of Members in the Central and Outer Regions of Galaxy
  Clusters}.
\newblock {\em {Astrophys. J.}} {\bf 2013}, {\em 768},~116.
\newblock {\url{https://doi.org/10.1088/0004-637X/768/2/116}}.

\bibitem[{Zou} et~al.(2024){Zou}, {Sui}, {Saintonge}, {Scholte}, {Moustakas},
  {Siudek}, {Dey}, {Juneau}, {Guo}, {Canning}, {Aguilar}, {Ahlen}, {Brooks},
  {Claybaugh}, {Dawson}, {de la Macorra}, {Doel}, {Forero-Romero}, {Gontcho A
  Gontcho}, {Honscheid}, {Landriau}, {Le Guillou}, {Manera}, {Meisner},
  {Miquel}, {Nie}, {Poppett}, {Rezaie}, {Rossi}, {Sanchez}, {Schubnell}, {Seo},
  {Tarl{\'e}}, {Zhou}, and {Zou}]{2024ApJ...961..173Z}
{Zou}, H.; {Sui}, J.; {Saintonge}, A.; {Scholte}, D.; {Moustakas}, J.;
  {Siudek}, M.; {Dey}, A.; {Juneau}, S.; {Guo}, W.; {Canning}, R.;  et~al.
\newblock {A Large Sample of Extremely Metal-poor Galaxies at z < 1 Identified
  from the DESI Early Data}.
\newblock {\em {Astrophys. J.}} {\bf 2024}, {\em 961},~173.
\newblock {\url{https://doi.org/10.3847/1538-4357/ad1409}}.

\bibitem[{Boquien} et~al.(2019){Boquien}, {Burgarella}, {Roehlly}, {Buat},
  {Ciesla}, {Corre}, {Inoue}, and {Salas}]{2019A&A...622A.103B}
{Boquien}, M.; {Burgarella}, D.; {Roehlly}, Y.; {Buat}, V.; {Ciesla}, L.;
  {Corre}, D.; {Inoue}, A.K.; {Salas}, H.
\newblock {CIGALE: A python Code Investigating GALaxy Emission}.
\newblock {\em {Astron. Astrophys.}} {\bf 2019}, {\em 622},~A103.
\newblock {\url{https://doi.org/10.1051/0004-6361/201834156}}.

\bibitem[{Bonato} et~al.(2017){Bonato}, {Negrello}, {Mancuso}, {De Zotti},
  {Ciliegi}, {Cai}, {Lapi}, {Massardi}, {Bonaldi}, {Sajina},
  {Smol{\v{c}}i{\'c}}, and {Schinnerer}]{Bonato2017}
{Bonato}, M.; {Negrello}, M.; {Mancuso}, C.; {De Zotti}, G.; {Ciliegi}, P.;
  {Cai}, Z.Y.; {Lapi}, A.; {Massardi}, M.; {Bonaldi}, A.; {Sajina}, A.;  et~al.
\newblock {Does the evolution of the radio luminosity function of star-forming
  galaxies match that of the star formation rate function?}
\newblock {\em Mon. Not. R. Astron. Soc.} {\bf 2017}, {\em
  469},~1912--1923.
\newblock {\url{https://doi.org/10.1093/mnras/stx974}}.

\bibitem[{Galluzzi} et~al.(2025){Galluzzi}, {Behiri}, {Giulietti}, and
  {Lapi}]{Galluzzi2025}
{Galluzzi}, V.; {Behiri}, M.; {Giulietti}, M.; {Lapi}, A.
\newblock {An Updated Repository of Sub-mJy Extragalactic Source-Count
  Measurements in the Radio Domain}.
\newblock {\em Galaxies} {\bf 2025}, {\em 13},~34.
\newblock {\url{https://doi.org/10.3390/galaxies13020034}}.

\bibitem[{Dow} and {White}(1995)]{1995ApJ...439..113D}
{Dow}, K.L.; {White}, S.D.M.
\newblock {ROSAT Observations of Coma Cluster Galaxies}.
\newblock {\em Astrophys. J.} {\bf 1995}, {\em 439},~113.
\newblock {\url{https://doi.org/10.1086/175156}}.

\bibitem[{Wetzel} et~al.(2012){Wetzel}, {Tinker}, and {Conroy}]{Wetzel2012}
{Wetzel}, A.R.; {Tinker}, J.L.; {Conroy}, C.
\newblock {Galaxy evolution in groups and clusters: Star formation rates, red
  sequence fractions and the persistent bimodality}.
\newblock {\em {Mon. Not. R. Astron. Soc.}} {\bf 2012}, {\em 424},~232--243.
\newblock {\url{https://doi.org/10.1111/j.1365-2966.2012.21188.x}}.

\bibitem[{Fabian}(2012)]{Fabian2012}
{Fabian}, A.C.
\newblock {Observational Evidence of Active Galactic Nuclei Feedback}.
\newblock {\em Annu. Rev. Astron Astrophys} {\bf 2012}, {\em
  50},~455--489.
\newblock {\url{https://doi.org/10.1146/annurev-astro-081811-125521}}.

\bibitem[{Tadhunter}(2016)]{Tadhunter2016}
{Tadhunter}, C.
\newblock {Radio AGN in the local universe: Unification, triggering and
  evolution}.
\newblock {\em Astron. Astrophys. Rev.} {\bf 2016}, {\em 24},~10.
\newblock {\url{https://doi.org/10.1007/s00159-016-0094-x}}.

\bibitem[{Chen} et~al.(2020){Chen}, {Sun}, {Yagi}, {Bravo-Alfaro}, {Brinks},
  {Kenney}, {Combes}, {Sivanandam}, {Jachym}, {Fossati}, {Gavazzi}, {Boselli},
  {Nulsen}, {Sarazin}, {Ge}, {Yoshida}, and {Roediger}]{2020MNRAS.496.4654C}
{Chen}, H.; {Sun}, M.; {Yagi}, M.; {Bravo-Alfaro}, H.; {Brinks}, E.; {Kenney},
  J.; {Combes}, F.; {Sivanandam}, S.; {Jachym}, P.; {Fossati}, M.;  et~al.
\newblock {The ram pressure stripped radio tails of galaxies in the Coma
  cluster}.
\newblock {\em {Mon. Not. R. Astron. Soc.}} {\bf 2020}, {\em 496},~4654--4673.
\newblock {\url{https://doi.org/10.1093/mnras/staa1868}}.

\bibitem[{Igo} et~al.(2024){Igo}, {Merloni}, {Hoang}, {Buchner}, {Liu},
  {Salvato}, {Arcodia}, {Bellstedt}, {Br{\"u}ggen}, {Croston}, {de Gasperin},
  {Georgakakis}, {Hardcastle}, {Nandra}, {Ni}, {Pasini}, {Shimwell}, and
  {Wolf}]{Igo2024}
{Igo}, Z.; {Merloni}, A.; {Hoang}, D.; {Buchner}, J.; {Liu}, T.; {Salvato}, M.;
  {Arcodia}, R.; {Bellstedt}, S.; {Br{\"u}ggen}, M.; {Croston}, J.H.;  et~al.
\newblock {The LOFAR - eFEDS survey: The incidence of radio and X-ray AGN and
  the disk-jet connection}.
\newblock {\em {Astron. Astrophys.}} {\bf 2024}, {\em 686},~A43.
\newblock {\url{https://doi.org/10.1051/0004-6361/202349069}}.

\bibitem[{Condon}(1992)]{Condon1992}
{Condon}, J.J.
\newblock {Radio emission from normal galaxies.}
\newblock {\em {Annu. Rev. Astron. Astrophys.}} {\bf 1992}, {\em 30},~575--611.
\newblock {\url{https://doi.org/10.1146/annurev.aa.30.090192.003043}}.

\bibitem[{G{\"u}rkan} et~al.(2018){G{\"u}rkan}, {Hardcastle}, {Smith}, {Best},
  {Bourne}, {Calistro-Rivera}, {Heald}, {Jarvis}, {Prandoni}, {R{\"o}ttgering},
  {Sabater}, {Shimwell}, {Tasse}, and {Williams}]{Gurkan2018}
{G{\"u}rkan}, G.; {Hardcastle}, M.J.; {Smith}, D.J.B.; {Best}, P.N.; {Bourne},
  N.; {Calistro-Rivera}, G.; {Heald}, G.; {Jarvis}, M.J.; {Prandoni}, I.;
  {R{\"o}ttgering}, H.J.A.;  et~al.
\newblock {LOFAR/H-ATLAS: The low-frequency radio luminosity-star formation
  rate relation}.
\newblock {\em Mon. Not. R. Astron. Soc.} {\bf 2018}, {\em
  475},~3010--3028.
\newblock {\url{https://doi.org/10.1093/mnras/sty016}}.

\bibitem[{Best} et~al.(2023){Best}, {Kondapally}, {Williams}, {Cochrane},
  {Duncan}, {Hale}, {Haskell}, {Ma{\l}ek}, {McCheyne}, {Smith}, {Wang},
  {Botteon}, {Bonato}, {Bondi}, {Calistro Rivera}, {Gao}, {G{\"u}rkan},
  {Hardcastle}, {Jarvis}, {Mingo}, {Miraghaei}, {Morabito}, {Nisbet},
  {Prandoni}, {R{\"o}ttgering}, {Sabater}, {Shimwell}, {Tasse}, and {van
  Weeren}]{Best2023}
{Best}, P.N.; {Kondapally}, R.; {Williams}, W.L.; {Cochrane}, R.K.; {Duncan},
  K.J.; {Hale}, C.L.; {Haskell}, P.; {Ma{\l}ek}, K.; {McCheyne}, I.; {Smith},
  D.J.B.;  et~al.
\newblock {The LOFAR Two-metre Sky Survey: Deep Fields data release 1. V.
  Survey description, source classifications, and host galaxy properties}.
\newblock {\em {Mon. Not. R. Astron. Soc.}} {\bf 2023}, {\em 523},~1729--1755.
\newblock {\url{https://doi.org/10.1093/mnras/stad1308}}.

\bibitem[{Gladders} and {Yee}(2000)]{Gladders2000}
{Gladders}, M.D.; {Yee}, H.K.C.
\newblock {A New Method For Galaxy Cluster Detection. I. The Algorithm}.
\newblock {\em {Astron. J.}} {\bf 2000}, {\em 120},~2148--2162.
\newblock {\url{https://doi.org/10.1086/301557}}.

\bibitem[{Sciarratta} et~al.(2019){Sciarratta}, {Chiosi}, {D'Onofrio}, and
  {Cariddi}]{Sciarratta2019}
{Sciarratta}, M.; {Chiosi}, C.; {D'Onofrio}, M.; {Cariddi}, S.
\newblock {Cosmological Interpretation of the Color-Magnitude Diagrams of
  Galaxy Clusters}.
\newblock {\em Astrophys. J.} {\bf 2019}, {\em 870},~70.
\newblock {\url{https://doi.org/10.3847/1538-4357/aaf00d}}.

\bibitem[{Rhee} et~al.(2017){Rhee}, {Smith}, {Choi}, {Yi}, {Jaff{\'e}},
  {Candlish}, and {S{\'a}nchez-J{\'a}nssen}]{Rhee2017}
{Rhee}, J.; {Smith}, R.; {Choi}, H.; {Yi}, S.K.; {Jaff{\'e}}, Y.; {Candlish},
  G.; {S{\'a}nchez-J{\'a}nssen}, R.
\newblock {Phase-space Analysis in the Group and Cluster Environment: Time
  Since Infall and Tidal Mass Loss}.
\newblock {\em {Astrophys. J.}} {\bf 2017}, {\em 843},~128.
\newblock {\url{https://doi.org/10.3847/1538-4357/aa6d6c}}.

\bibitem[{Pasquali} et~al.(2019){Pasquali}, {Smith}, {Gallazzi}, {De Lucia},
  {Zibetti}, {Hirschmann}, and {Yi}]{Pasquali2019}
{Pasquali}, A.; {Smith}, R.; {Gallazzi}, A.; {De Lucia}, G.; {Zibetti}, S.;
  {Hirschmann}, M.; {Yi}, S.K.
\newblock {Physical properties of SDSS satellite galaxies in projected phase
  space}.
\newblock {\em {Mon. Not. R. Astron. Soc.}} {\bf 2019}, {\em 484},~1702--1723.
\newblock {\url{https://doi.org/10.1093/mnras/sty3530}}.

\bibitem[{Dou} and {Yu}(2025)]{2025A&A...699A..95D}
{Dou}, H.; {Yu}, H.
\newblock {Estimating infall times of galaxies around clusters: Working to
  achieve accurate results based on observational data}.
\newblock {\em {Astron. Astrophys.}} {\bf 2025}, {\em 699},~A95.
\newblock {\url{https://doi.org/10.1051/0004-6361/202555240}}.

\bibitem[{Healy} et~al.(2021){Healy}, {Blyth}, {Verheijen}, {Hess}, {Serra},
  {van der Hulst}, {Jarrett}, {Yim}, and {J{\'o}zsa}]{Healy2021}
{Healy}, J.; {Blyth}, S.L.; {Verheijen}, M.A.W.; {Hess}, K.M.; {Serra}, P.;
  {van der Hulst}, J.M.; {Jarrett}, T.H.; {Yim}, K.; {J{\'o}zsa}, G.I.G.
\newblock {H I content in Coma cluster substructure}.
\newblock {\em Astron. Astrophys.} {\bf 2021}, {\em 650},~A76.
\newblock {\url{https://doi.org/10.1051/0004-6361/202038738}}.

\bibitem[{Lyskova} et~al.(2019){Lyskova}, {Churazov}, {Zhang}, {Forman},
  {Jones}, {Dolag}, {Roediger}, and {Sheardown}]{Lyskova2019}
{Lyskova}, N.; {Churazov}, E.; {Zhang}, C.; {Forman}, W.; {Jones}, C.; {Dolag},
  K.; {Roediger}, E.; {Sheardown}, A.
\newblock {Close-up view of an ongoing merger between the NGC 4839 group and
  the Coma cluster---A post-merger scenario}.
\newblock {\em {Mon. Not. R. Astron. Soc.}} {\bf 2019}, {\em 485},~2922--2934.
\newblock {\url{https://doi.org/10.1093/mnras/stz597}}.

\end{thebibliography}
\end{document}